\documentclass[a4paper,11pt]{article}

\usepackage{graphicx}
\usepackage{float}
\usepackage{wrapfig}
\usepackage{xcolor}
\usepackage[T1]{fontenc}
\usepackage{epsfig}
\usepackage{color}
\usepackage{amsmath,jheppub,mathtools}
\usepackage{subfloat}
\usepackage{amsfonts}
\usepackage{braket}
\usepackage{cleveref}
\usepackage{epstopdf}
\usepackage{caption}
\usepackage{subcaption}
\usepackage{enumitem}
\usepackage[numbers]{natbib}
\usepackage[titletoc,toc,title]{appendix}
\usepackage{hyperref}
\hypersetup{
	colorlinks=true,
	linkcolor=blue,
	filecolor=red,      
	urlcolor=blue,
	citecolor=blue
} 

\title{\boldmath Krylov Complexity in $2d$ CFTs with SL$(2,\mathbb{R})$ deformed Hamiltonians}

\author[a]{Vinay Malvimat,}
\author[b]{Somnath Porey,}
\author[b]{Baishali Roy}
\affiliation[a]{Asia Pacific Center for Theoretical Physics, 77 Cheongam-ro, Nam-gu, Pohang-si, Gyeongsangbuk-do, 37673, Korea.}
\affiliation[b]{Ramakrishna Mission Vivekananda Educational and Research Institute, Belur Math, Howrah-711202, West Bengal, India}
\affiliation[c]{}

\emailAdd{vinay.malvimat@apctp.org}
\emailAdd{somnathhimu00@gm.rkmvu.ac.in}
\emailAdd{baishali.roy025@gm.rkmvu.ac.in}

\date{}

\abstract{ In this study, we analyze Krylov Complexity in two-dimensional conformal field theories subjected to deformed SL$(2,\mathbb{R})$ Hamiltonians. In the Vacuum state, we find that the K-complexity exhibits a universal phase structure. The phase structure involves the K-complexity exhibiting an oscillatory behaviour in the non-heating phase, which contrasts with the exponential growth observed in the heating phase, while it displays polynomial growth at the phase boundary. Furthermore, we extend our analysis to compute the K-complexity of a light operator in excited states, considering both large-c CFT and free field theory. In the free field theory, we find a state-independent phase structure of K-complexity. However, in the large-c CFT, the behavior varies, with the K-Complexity once again displaying exponential growth in the heating phase and polynomial growth at the phase boundary. Notably, the precise exponent governing this growth depends on the heaviness of the state under examination. In the non-heating phase, we observe a transition in K-complexity behavior from oscillatory to exponential growth, akin to findings in \cite{Kundu:2023hbk}, as it represents a special case within the non-heating phase.}

\begin{document} 
	\maketitle
	
	\flushbottom


			\section{Introduction}
	
  Quantum Chaos has emerged as a frontier field of intense research exploration captivating the attention of both high energy as well as quantum many-body physicists. Several interesting probes including the level spacing statistics, the exponential growth of the Out-of -Time Order Correlators (OTOC) at late times, dip-ramp-plateau behaviour of the Spectral Form Factor(SFF) have provided illuminating insights into the understanding of thermalization, ergodicity, and chaos in various quantum many-body systems \cite{Srednicki:1994mfb,Gogolin:2015gts, DAlessio:2015qtq}. 
 In \cite{Roberts:2014ifa}, it was shown that the four-point OTOCs in the thermal state can distinguish between chaotic (eg. large-c) and integrable (eg. Ising model) 2D CFTs. Subsequently, other studies have provided evidence for this even in the non-equilibrium setup of globally quenched \cite{Das:2021qsd} and driven \cite{Das:2022jrr} CFTs.



 
The interesting interplay between the onset of quantum chaos and the growth of complexity is also a fascinating area of intense investigation. An interesting measure that characterizes this growth is the so-called \textit{circuit complexity}.
It is defined in quantum information theory as the minimum number of unitary gates in a quantum circuit required to reach a specific state from a reference state. Interestingly, the growth of complexity in holography is expected to be dual to the growth of the volume of the black hole interior in the dual bulk gravitational theory \cite{Stanford:2014jda, Susskind:2018pmk, Susskind:2018tei,Chattopadhyay:2023fob}.  While these studies have led to many  interesting insights, 
several questions regarding circuit complexity remain. These include defining the precise notion of distance between states in quantum field theories and finding the exact bulk dual of circuit complexity in holography (See \cite{Jefferson:2017sdb,Chapman:2017rqy,Chagnet:2021uvi,Belin:2021bga} and the references therein for many interesting discussions).

 Recently, a new notion of complexity known as Krylov complexity characterizing the growth of a local operator in operator Hilbert space (Krylov space) has received considerable attention.  This quantity was introduced in \cite{Parker:2018yvk}, where the authors proposed that the Lanczos coefficients associated with Krylov space grow linearly (maximally) in a system that exhibits quantum chaos. This is known as the operator growth hypothesis. Following this development, Krylov complexity and the associated Lanczos coefficients have been examined in varied systems including quantum and conformal field theories \cite{Dymarsky:2019elm,Avdoshkin:2019trj,Barbon:2019wsy,Dymarsky:2021bjq,Caputa:2021sib,Caputa:2021ori,Avdoshkin:2022xuw,Camargo:2022rnt,Khetrapal:2022dzy, Rabinovici:2022beu}, random matrix theories \cite{Kar:2021nbm,Erdmenger:2023wjg,	Tang:2023ocr}, holography \cite{Rabinovici:2023yex,Anegawa:2024wov,Basu:2024tgg}, open quantum systems \cite{Bhattacharjee:2022lzy}, Matrix models \cite{Iizuka:2023pov,Iizuka:2023fba}, Scar states \cite{Bhattacharjee:2022qjw},  driven quantum systems \cite{Nizami:2023dkf} among others\footnote{A related notion known as spread complexity has also been explored extensively for various quantum systems \cite{Balasubramanian:2022tpr,Caputa:2022eye,Caputa:2022yju,Afrasiar:2022efk,Chattopadhyay:2023fob} }. Furthermore, the Krylov complexity is conjectured to provide an upper bound for the Lyapunov exponent obtained from OTOC \cite{Parker:2018yvk}
 \begin{align}
    \lambda_L\leq\lambda_K\leq \frac{2\pi}{\beta}
 \end{align}
	where $\lambda_L$ is the Lypunov exponent, $\lambda_K$ the Krylov exponent and $\beta$ is the inverse temperature.

In \cite{Dymarsky:2021bjq}, the authors computed K-Complexity in free scalar and fermionic quantum field theories and 2d CFTs utilizing the Wightman inner product for Krylov space. They demonstrated that the Lanczos coefficents grow linearly and that the K-Complexity grows exponentially in all cases. Since for 2D CFTs, the auto-correlation function is related to the thermal two-point function, the exponential behaviour of K-Complexity turns out to be universal. In \cite{Kundu:2023hbk}, the authors addressed this intriguing issue by choosing a state-dependent innerproduct. In 2d CFTs, they obtained the K-Complexity for large-c CFTs and Ising model for various states. They showed that while the Lanczos coefficients grow linearly for large-n, the K-complexity exhibits a transition from an oscillating to an exponentially growing behaviour beyond a critical dimension of the primary operator, creating the state by acting on the vacuum.
However, in two-dimensional free field theory and Ising model the K-complexity always showed an oscillatory behaviour, independent of the dimension of the primary operator.


The main focus of this draft is to examine the dynamics of K-complexity in out-of-equilibrium systems. We use quantum quenches in  (1+1)D CFT defined on a circle as our basic framework, which provides a tractable yet insightful setup for this investigation. Different types of quench protocols have been studied in the literature, in the context of 2D CFTs. For example, one such extensively studied protocol is that of global quench setup \cite{cardy}, in which the system is initially prepared in the regulated boundary state of a CFT and is then evolved with the usual CFT Hamiltonian. 
In this article, we will consider yet another kind of quench, where the system is prepared in some eigenstate of the usual CFT Hamiltonian and is then evolved with a deformed CFT Hamiltonian. 

The usual CFT Hamiltonian of 2D CFT on a ring is built out of only the Virasoro modes $L_0$ and $\bar{L}_0$, which generate scaling in the radial quantization. One can, however, consider a situation where Hamiltonian consists of other Virasoro generators. Specifically one can study systems under SL$_q[2,\mathbb{R}]$ deformations where the Hamiltonian is made out of the linear combination of generators of sl$_2$ subalgebra of the Virasoro algebra rendering the unitary evolution operators to be $su(1,1)$ valued that generate Möbius transformations in the q-sheet Riemann Surface\cite{Fan:2020orx}. A well-known fact about these classes of deformed CFTs is that there are three distinct types of evolution operators, depending on the sign of the Casimir of the underlying algebra, which results in three distinct classes of dynamics - hyperbolic (heating phase), elliptic (non-heating phase) and parabolic (phase boundary)\cite{Wen:2018agb, Wen:2021mlv, Fan:2020orx, Wen:2020wee,Wen:2022pyj}\footnote{  Understanding the spectrum of the Hamiltonians in each of these phases is an interesting problem, \textit{see } \cite{Ishibashi:2015jba}}. 
In particular, it was shown in \cite{Das:2022jrr} that OTOC in large-c driven CFTs shows qualitatively different behavior in each of the three phases. 
While in the heating phase, it shows the characteristic Lyapunov behaviour, in the non-heating phase and phase boundary it shows oscillatory and polynomial behaviour respectively.
In light of these results, it would be interesting to understand whether K-complexity can distinguish between different phases in such systems and also whether it can demarcate between large-c and integrable CFTs. In this draft, we have tried to address these questions. Our key findings are as follows:

\begin{itemize}
    \item  In the vacuum state, the K-complexity is determined from a two-point function and it shows exponential, oscillatory, and polynomial behaviour respectively in the heating phase, non-heating phase, and at the phase boundary. 
    
    \item For large-c CFTs, in the non-heating phase, behaviour of K-complexity in the excited state shows qualitatively different behaviour for states with conformal dimension ($\Delta$) above and below a certain threshold value($\Delta^*=\frac{c}{12})$. The temporal dependence is exponential and oscillatory respectively above and below $\Delta^*$. For, the free massless scalar field theory, however, the K-complexity shows only oscillatory behaviour for all excited states. 
    

     \item On the other hand, in the heating phase, and at the phase boundary, the K-complexity shows the same behavior in all excited states both for large-c CFT as well as for free massless scalar field theory.  In particular, it shows exponential and polynomial growth respectively for the heating phase and phase boundary dynamics.
\end{itemize}


The rest of the draft is organized as follows. We begin with a brief review of K-complexity in section \ref{kcRvw}.  In section \ref{mainRes}, we compute the auto-correlation functions and examine the behaviour of K-complexity in the deformed CFT setup. In \ref{vacuumUniiv}, we set up the computation of the K-complexity in the CFT ground state. This is followed by the computation of the same in an excited state in \ref{extd}. 
   In the excited state, we carry out this computation in large-c  (\ref{extdLargeC}) as well as in the free massless scalar field theory(\ref{extdscalar}). We wrap up by discussing our results and suggesting potential future directions in section \ref{dissc}. To enhance the overall coherence and completeness of the draft, we also include two appendices at the end. In appendix \ref{appndxA}, we have reviewed the details of deformed CFTs, and in appendix \ref{appndxB}, we have discussed the details of the numerical technique utilized in this paper.

				\section{Brief review of K-complexity in CFTs}\label{kcRvw}
 The time evolution of a local operator $\mathcal{O}$ in the Heisenberg picture can be expressed as a sum of nested commutators of the operator with the Hamiltonian $H$ 
			\begin{equation}
				\mathcal{O}(t)=e^{iHt}\mathcal{O}(0)e^{-i Ht}=\sum_{k=0}^{\infty} \frac{(-it)^k}{k!} [H,[H\dots[H,\mathcal{O}(0)]\dots]]
			\end{equation}
			By defining the \textbf{Liouvillian operator}  $\mathcal{L}:=[H,\cdot]$, the above expression can be 
  re-expressed as follows
			\begin{equation*}
				\mathcal{O}(t)=e^{i \mathcal{L} t} \mathcal{O}(0):=\sum_{k=0}^{\infty} \frac{(-it)^k}{k!}\tilde{\mathcal{O}_k}
			\end{equation*}
			where, $\tilde{\mathcal{O}_k}=\mathcal{L}^k\mathcal{O}(0)$ and $\mathcal{L}^k$ are nested commutators $[H,[H\dots[H,\mathcal{O}]\dots]]$  with the Hamiltonian. The set of operators $\{ \tilde{\mathcal{O}_k}\}$  span a vector space known as the \textbf{Krylov space} associated with the operator $\mathcal{O}$. One can construct an orthonormal basis in this space, known as the \textbf{Krylov basis} $\{ \mathcal{O}_k\}$, through the standard Gram-Schmidt orthogonalization which is implemented through an iterative procedure known as \textit{Lanczos algorithm}. The Krylov space has to be endowed with a valid inner product. The standard choice for this is the Wightmann inner product.
			 However, as pointed out in \cite{Dymarsky:2019elm} various other choices can be made. A class of such valid inner products is of the form
				\begin{equation*}
				(\mathcal{O}_1|\mathcal{O}_2)=Tr(\rho_{1} \mathcal{O}_1^{\dagger}\rho_{2}\mathcal{O}_2)
			\end{equation*}
			where $\rho_1$ and $\rho_2$ are positive Hermitian operators that commute with $H$.	A particular choice of $\rho_1=\rho_{\psi}$, which is the density matrix associated with a pure eigen state $\ket{\psi}$  and $\rho_2=I$ was chosen in \cite{Kundu:2023hbk}
			\begin{equation*}
				(\mathcal{O}_1|\mathcal{O}_2)= \frac{Tr(\rho_{\psi} \mathcal{O}_1^{\dagger}\mathcal{O}_2)}{Tr \rho_{\psi}}
			\end{equation*}
		The orthonormal basis constructed utilizing \textit{Lanczos Algorithm}) is given as
			\begin{align}
				|\mathcal{P}_n):&=\mathcal{L}|\mathcal{O}_{n-1})+b_{n-1} |\mathcal{O}_{n-2})- a_{n-1} |\mathcal{O}_{n-1})\\\nonumber
				|\mathcal{O}_n):&=b_n^{-1}|\mathcal{P}_n)\\\nonumber
				b_n:&=\sqrt{ (\mathcal{P}_n|\mathcal{P}_n)}
			\end{align}
			The coefficients $a_n$ and $b_n$ so generated are called the Lanczos coefficients. They can be expressed in terms of the matrix elements of the $\mathcal{L}$ as,
			\begin{equation}
				a_n:= (\mathcal{O}_n|\mathcal{L}|\mathcal{O}_n)\ \  \ \ \text{and} \ \ \  b_n:=(\mathcal{O}_n|\mathcal{L}|\mathcal{O}_{n-1}).
			\end{equation}
			The Liouvillian can be expressed as a tridiagonal matrix in this basis:
			\begin{equation}
				\mathcal{L}_{mn}=\begin{pmatrix}
					a_0 & b_1 & 0& 0 &0 & 0 & \dots & 0\\
					b_1 & a_1 &  b_2& 0 &0 &0 &\dots & 0\\
					0 & b_2& a_2& b_3& 0 &0 &\dots & 0\\
					0 & 0 & b_3& a_3& b_4& 0 & \dots & 0\\
					\vdots& \vdots & \vdots & \ddots & \ddots &\ddots &\dots &0\\
					\vdots & \vdots& \vdots & \vdots & \ddots & \ddots &\ddots  &0\\
					\vdots & \vdots& \vdots & \vdots & \vdots & \ddots &\ddots  &b_m\\
					0& 0 & 0 & 0 & 0 &\dots& b_m  &a_m
				\end{pmatrix}
			\end{equation}
			Having defined a Krylov space with a valid inner product, the operator $\mathcal{O}(t)$ at any moment in time  can be expanded in terms of the Krylov basis elements $|\mathcal{O}_n)$: 
			\begin{equation*}
				|\mathcal{O}(t)\ )= \sum_{n} i^n \phi_n(t)|\mathcal{O}_n)
			\end{equation*}
			where $\phi_n(t)$’s are complex coefficients. The Heisenberg equation of motion essentially reduces to  a recursion relation of these $\phi_n$'s, as given below
		    	\begin{align}\label{phin}
			    \partial_t \phi_n(t)=b_n \phi_{n-1}(t)-b_{n+1}\phi_{n+1}(t)+i a_n \phi_n(t)
	     	   \end{align}
			with the boundary condition $\phi_{-1}(t)=0, b_0=0$ and $\phi_n(0)=\delta_{n,0}$. $\phi_0$ 
			in our case is going to be the unequal-time correlation (or auto-correlation) function $(\mathcal{O}(0)|\mathcal{O}(t))$ in some state $\psi$.  
			$$ C_{\psi}(t)=(\mathcal{O}(0)|\mathcal{O}(t))_{\psi}=\phi_0(t)=\frac{Tr(\rho_{\psi} \mathcal{O}(t)\mathcal{O}(0))}{Tr \rho_{\psi}}$$
			The Krylov-complexity is then defined as,
			$$K_{\mathcal{O}}(t)=(\mathcal{O}(t)|n|\mathcal{O}(t))=\sum_n n |\phi_{n}(t)|^2$$

			\section{From auto-correlation function to K-complexity}\label{mainRes}
	We consider a two-dimensional CFT on a cylinder whose circumference is of length L. The system is described by the coordinates $\omega=\tau_c+ i x_c$ and $\bar{\omega}=\tau_c-i x_c$ where $\tau_c$ is the Euclidean time and $x_c$ is the compact spatial direction. The standard  CFT Hamiltonian in terms of Virasoro generator is given by $H=L_0+\bar{L}_0$\footnote{Here, we ignore the constant casimir term $-\frac{c}{24}$, since it does not play any crucial role in our computation.}. In the present article, we will instead focus our attention on the quantum system which is evolved with a  Hamiltonian that is deformed by the presence of other Virasoro generators as described below,	
			\begin{align}\label{defh}
				H^{cyl}_d=\alpha(L_{0}+\bar{L}_{0})+\beta (L_{q+}+\bar{L}_{q+})+ \gamma (L_{q-}-\bar{L}_{q-})
			\end{align}
			with, $L_{q+}=\frac{L_{q}+L_{-q}}{2}$, $L_{q-}=\frac{L_{q}-L_{-q}}{2 i}$ and $\alpha,\ \beta $ and $\gamma$ are some real numbers. 
			The dynamics of such a Hamiltonian can be classified into three distinct phases depending on the sign of the parameter $\delta=-\alpha^2+\beta^2+\gamma^2$ \cite{Wen:2020wee}.
CFTs with such deformed Hamiltonians have been investigated in great detail in \cite{Wen:2018agb, Wen:2021mlv, Fan:2020orx, Wen:2020wee,Wen:2022pyj}. We direct the reader to a brief review of the same in Appendix \ref{appndxA}.


			To compute the Krylov complexity of an operator evolving with the deformed Hamiltonian, we consider the generic form of the auto-correlation function $C_{\phi}(\tau)$: 
			\begin{equation}\label{autocor}
				C_{\phi}(\tau)=\frac{\langle \phi|\mathcal{O}(\tau)\mathcal{O}(0)|\phi\rangle}{\langle \phi|\phi\rangle}
			\end{equation}
			where, the operator $\mathcal{O}(\tau)$ in the Heisenberg picture is given by,
			\begin{align}
			\mathcal{O}(\tau)=e^{H^{cyl}_d \tau}\mathcal{O}(0)e^{- H^{cyl}_d \tau}
			\end{align} 
			where $ H^{cyl}_d $ is the deformed Hamiltonian on the ring defined in \eqref{defh} and $\tau$ is the corresponding Euclidean time generated by $ H^{cyl}_d $. 	
			In the following, we aim to analyze the K-complexity of such systems for different cases. 
			
		\subsection{Vacuum state of a 2d CFT}\label{vacuumUniiv}
			
			In this section, we consider the situation, when the auto-correlation function is evaluated in the vacuum state of the CFT. 
   For computational convenience, we work with the deformed Hamiltonian with $q=1$ in \eqref{defh} which is as follows 
			\begin{align}\label{defh2}
				H^{cyl}_d=\alpha(L_{0}+\bar{L}_{0})+\beta (L_{+}+\bar{L}_{+})+ \gamma (L_{-}-\bar{L}_{-})
			\end{align}
			where, $L_{+}=\frac{L_{1}+L_{-1}}{2}$, $L_{-}=\frac{L_{1}-L_{-1}}{2 i}$. Since, the above deformed Hamiltonian  is made out of the global Virasoro generators $L_0,L_1,L_{-1}$ and their corresponding complex conjugates, they generate a M\"{o}bius transformation in the complex plane:
			\begin{align*}
				z'_{i}=\frac{a z_{i}+b}{c z_{i}+d}
			\end{align*}
			where, $z_i$ are complex coordinates on the plane and $z'_i$ are the corresponding coordinates after the Möbius transformation. Note that the vacuum state (ground state of $H_0=L_0+\bar{L}_0$) is invariant under the action of these generators and hence, only the unequal time correlation functions have access to the dynamics of the drive.  To compute the auto-correlation function, we first map the correlator from cylinder($\omega, \bar{\omega}$) to the plane($z, \bar{z}$) via the coordinate transformations \footnote{We have taken $L=2 \pi$ and in the rest of the paper we follow this convention.},
			\begin{align}
				z=e^{\omega},\hskip 0.5cm \bar{z}=e^{\bar{\omega}}.
			\end{align}
		   We will also consider the operators $1\ \text{and}\ 2$ to be initially placed at coordinate points $x_{c\ 1,2}=\tau_{c\ 1,2}=0$ on the cylinder.

 The auto-correlation function is given by,
			\begin{align}\label{dacf1}
				\langle 0|\mathcal{O}(\tau)\mathcal{O}(0)|0\rangle_{cyl}&= |z_1|^{\Delta} |z_2|^{\Delta} \bigg|\frac{\partial z_{1\tau}}{z_1}\bigg|^{\Delta}\langle 0|\mathcal{O}(z_{1\tau}, \bar{z}_{1\tau})\mathcal{O}(z_2,\bar{z}_2)|0\rangle_{pl}\\\nonumber
				&= \frac{1}{|c_{\tau}+d_{\tau}|^{2 \Delta}}\langle 0|\mathcal{O}\bigg(\frac{a_{\tau}+b_{\tau}}{c_{\tau}+d_{\tau}},\frac{\bar{a}_{\tau}+\bar{b}_{\tau}}{\bar{c}_{\tau}+\bar{d}_{\tau}}\bigg)\mathcal{O}(0)|0\rangle_{pl}\\\nonumber
				&= \frac{1}{|a_{\tau}+b_{\tau}-c_{\tau}-d_{\tau}|^{2 \Delta}}
			\end{align}
			
			where, in the second line of the equation, we write\footnote{The subscript $\tau$ in $a_{\tau}, b_{\tau}, c_{\tau}, d_{\tau}$ indicates that the parameters of the Möbius transformation vary with time, as elaborated in Appendix A.} $z_{1\tau}=g(z_1,\tau)=\frac{a_{\tau} z_1+b_{\tau}}{c_{\tau} z_1+d_{\tau}}$ and $\bar{z}_{1\tau}=\bar{g}(\bar{z}_1,\tau)=\frac{\bar{a}_{\tau} \bar{z}_1+\bar{b}_{\tau}}{\bar{c_{\tau}} \bar{z}_1+\bar{d}_{\tau}}$. $\Delta$ in $\eqref{dacf1}$ is the conformal dimension of the operator $\mathcal(O)$, and  $\Delta=2 h$ since we are considering scalar operators here. Now, following the earlier discussion, depending on the different values of the parameter $\delta$ one can distinguish the three phases and correspondingly one can also find three different expressions for the $(a_{\tau}, b_{\tau}, c_{\tau}, d_{\tau})$ for the three different phases. We discuss the details of this derivation in appendix \ref{appndxA}.
   

			 Using the expressions for $(a_{\tau}, b_{\tau}, c_{\tau}, d_{\tau})$ from \eqref{HPab}, \eqref{NHPab} and \eqref{PBab}, one can write down the analytic expression for the auto-correlation functions in the heating phase, non-heating phase and phase boundary respectively. Once the auto-correlation functions are known, one can find out the Lanczos Coefficients in each phase using the Toda-Chain (or Hirota’s bilinear) technique \cite{Dymarsky:2019elm}. As a first step to this technique, one uses the autocorrelation function $C_{\phi}$ to initialize the toda-function $\nu_0$ and solve the Toda Chain recursion relation
    \begin{equation}\label{taunrec}
        \nu_n \ddot{\nu}_n-\dot{\nu}^2_n=\nu_{n-1}\nu_{n+1}
    \end{equation}
    with $\nu_{-1}=0$ and $\nu_0=C_{\phi}(\tau)$, 
    to find an expression for toda-function after nth iteration i.e. $\nu_n$. This then allows us to compute the Lanczos coefficients using \eqref{taunrec} and its derivatives\footnote{A more detailed review of this technique can be found in the appendix of \cite{Kundu:2023hbk}}. The knowledge of $a_n\ \text{and}\ b_n$ then facilitates one to compute the normalized $\phi_n$ using \eqref{phin}. 
    
    Below we write down the analytic expressions for the auto-correlation function, Lanczos coefficients, and normalized $\phi_n$\footnote{After analytically continuing to real-time t by $\tau \rightarrow \tau_0+ it$ } in the three phases. \\

 \vskip 0.2cm

 \underline{\textbf{Non-heating Phase ($\delta<0$):}} \\

In this case,  substitution of \eqref{HPab} in \eqref{dacf1} leads to the auto-correlation function,
			\begin{align}{\label{twoptnhp}}
				C_{NHP}(\tau)&=\left( \frac{2 (\alpha +\beta ) \sinh \left(\frac{\sqrt{|\delta|}  \tau}{2}\right)}{\sqrt|\delta| }\right)^{-2 \Delta}
            \end{align}
Then as described earlier, using the Toda chain technique, the Lanczos coefficients can be obtained as,
               \begin{align}
                a_n &=-\frac{2 (\Delta + n) \sqrt{|\delta|}}{tanh(\sqrt{|\delta|} \tau_0)} \\
				b_n &=\frac{\sqrt{-n + 2 \Delta n + n^2} \sqrt{|\delta|}}{sinh (\sqrt{|\delta|}\tau_0)}
              \end{align}
Here, we notice that the Lanczos coefficients grow linearly with n which is expected in such computations. Using $a_n$ and $b_n$, we obtain the normalized wave function in this case as,
              \begin{align}
				\phi_n (t) &= \sqrt{\frac{\Gamma (2 \Delta+n)}{\Gamma (2 \Delta) \Gamma (n+1)}} \sin ^n(\sqrt{|\delta|}t) csch ^{n+2 \Delta}(\sqrt{|\delta|} (\tau_0+i t)) \left(-\frac{\sqrt{|\delta|}}{\alpha +\beta }\right)^{2 \Delta}
                \end{align}
                
wave function is oscillatory in time and we will see that the K-complexity for this case will turn out to be oscillatory as well (\ref{KNH}).
		\vskip 0.5cm

 \underline{\textbf{Heating Phase ($\delta>0)$:} }\\

 In this phase,  substitution of \eqref{NHPab} in \eqref{dacf1} leads to the auto-correlation function,

   \begin{align}{\label{twopthp}}
				C_{HP}(\tau)=\left( \frac{ 2 (\alpha +\beta ) \sin \left(\frac{\sqrt{|\delta|}  \tau }{2}\right)}{\sqrt{|\delta|} }\right)^{-2 \Delta}
               \end{align}
This leads to the corresponding Lanczos coefficients (that are linear functions in n), as follows:                
                \begin{align}    
                a_n &=-\frac{2 (\Delta + n) \sqrt{|\delta|}}{tan(\sqrt{|\delta|} \tau_0)} \\
				b_n &=\frac{\sqrt{-n + 2 \Delta n + n^2} \sqrt{|\delta|}}{sin (\sqrt{|\delta|}\tau_0)}
    \end{align}
    The analytically continued normalized wave function, in this case, becomes a hyperbolic function in t,  
    \begin{align}
        \phi_n (t) &= \sqrt{\frac{\Gamma (2 \Delta+n)}{\Gamma (2 \Delta) \Gamma (n+1)}} \sinh ^n(\sqrt{|\delta|}t) \csc ^{n+2 \Delta}(\sqrt{|\delta|} (\tau_0+i t)) \left(-\frac{\sqrt{|\delta|}}{\alpha +\beta }\right)^{2 \Delta}
			\end{align}
			
		and therefore, leads to the exponential k-complexity in the late time limit.
			
			\vskip 0.5cm

		\underline{\textbf{Phase Boundary ($\delta=0$):} }\\

	Upon substitution of \eqref{PBab} in \eqref{dacf1}, the one finds the correlation function for this phase to be

   \begin{align}{\label{twoptpb}}
				C_{PB}(\tau)&=(\tau  (\alpha +\beta ))^{-2 \Delta}
    \end{align}
    which leads to corresponding Lanczos coefficients,
    \begin{align}
    a_n &=-\frac{2 (\Delta + n)}{\tau_0} \\
				b_n &=\frac{\sqrt{-n + 2 \Delta n + n^2}}{\tau_0}
    \end{align}
			and subsequently, to the analytically continued normalized wave function
\begin{align}
	\phi_n (t) &= (- \tau_0)^{4 h} \frac{t^n}{(\tau_0+i t)^{2 \Delta+n}} \sqrt{\frac{\Gamma (2 \Delta+n)}{\Gamma (2 \Delta) \Gamma (n+1)}}. 
			\end{align}

   Notice that these expressions of the auto-correlation functions do not involve the parameter $\gamma$ in any phase. It is also particularly relevant for the heating phase because one possible choice of parameters for the heating phase is when $\alpha$ and $\beta$ are both zero but $\gamma$ is non-zero since this choice will also yield $\delta>0$. However, it seems from \eqref{twopthp} that the auto-correlation function diverges when $\alpha+\beta=0$. This issue gets resolved once we note that for this particular choice of $\alpha,\beta,\gamma$, initial position $x = 0$ corresponds to a fixed point of the problem, where z equals 1. In practical terms, this implies that under the deformed Hamiltonian evolution, the position at $x = 0$ remains unchanged. As a result, two of the operators essentially stay in the same position, leading to the observed divergence in the correlator.

			
			
			Note that, in $\phi_n(t)$, we have analytically continued the Euclidean time $\tau$ to $\tau_0+i t$. We also find that the coefficients $a_n$ and $b_n$ can be written in terms of $\tau_0$. 
			one can then easily find the exact analytic expression for each phase as given below 
			\begin{align}
				K_{HP}&=\frac{2 \Delta}{\sin^2\bigg(\frac{\sqrt{|\delta|} \tau_0}{2}\bigg)} \sinh^2\bigg(\frac{\sqrt{|\delta|} t}{2}\bigg) \ \ \  \ \ \ \ \ \ \ \textbf{(Heating Phase)}\\
				K_{NHP}&=\frac{2 \Delta}{\sinh^2\bigg(\frac{\sqrt{|\delta|} \tau_0}{2}\bigg)} \sin^2\bigg(\frac{\sqrt{|\delta|} t}{2}\bigg) \ \ \label{KNH} \ \ \ \ \ \ \textbf{(Non-heating Phase)}\\
				K_{PB}&= \frac {2 \Delta }{\tau_0^2} t^2 \ \ \ \ \  \ \ \ \ \ \ \ \ \ \hskip 3.3cm \ \ \textbf{(Phase Boundary)}
			\end{align}
			
			Surprisingly, we find that the behaviour of K-complexity for the heating and the non-heating phases that we have found in this section is similar to the behaviour of the two-point function in the generic state above and below the black hole threshold respectively found in \cite{Kundu:2023hbk}. Next, we study the behaviour of the K-complexity for deformed CFTs in a generic state. 

			\subsection{Excited state of a 2d CFT}\label{extd}
			
			In this section, we define the auto-correlation function $C_{\phi}(\tau)$ in an excited state, following \cite{Kundu:2023hbk} as, 
			\begin{equation}\label{autocor}
				C_{\psi}(\tau)=\frac{\langle \psi_{out}|\mathcal{O}_{L}(\tau)\mathcal{O}_{L}(0)|\psi_{in}\rangle}{\langle \psi_{out}|\psi_{in}\rangle}
			\end{equation}
			where the $in$ and $out$ states are defined as,
			$|\psi_{in}\rangle=\lim\limits_{T\to \infty }\mathcal{O}_{H}(T)|0\rangle$ and $\langle\psi_{out}|= \lim\limits_{T\to \infty} \langle0| \mathcal{O}_{H}(T) $ and $|0\rangle$ is the ground state of the 2d CFT on a circle \footnote{This is the ground state of the deformed CFT as well. Since the deformed Hamiltonian is made out of global conformal generators only.}.
			Noting that, $\mathcal{O}_{L}(\tau)=e^{H^{cyl}_d \tau}\mathcal{O}_{L}(0)e^{- H^{cyl}_d \tau}$, we can write (\ref{autocor}) explicitly as:
			\begin{equation}\label{autocor1}
				C_{\psi}(\tau)=\lim_{\substack{\omega_{1}\to \infty \\ \omega_{4}\to 0}}\frac{\langle 0|\mathcal{O}_{H}(\omega_{1},\bar{\omega_{1}})e^{H^{cyl} \tau}\mathcal{O}_{L}(\omega_{2},\bar{\omega_{2}})(0)e^{- H^{cyl} \tau}\mathcal{O}_{L}(\omega_{3},\bar{\omega_{3}})\mathcal{O}_{H}(\omega_{4},\bar{\omega_{4}})|0\rangle}{\langle 0| \mathcal{O}_{H}(\omega_{1},\bar{\omega_{1}})\mathcal{O}_{H}(\omega_{4},\bar{\omega_{4}})|0\rangle}
			\end{equation}

			Under the above coordinate transformations, the correlation function transforms as,
			\begin{equation}\label{fourpt}
				C_{\psi}(\tau)= \left|\frac{dz'_{2}}{dz_{2}}\right|^{ \Delta_{L}}\left|z_{14}\right|^{2 \Delta_{H}}\left|z_{2} z_{3}\right|^{2 \Delta_{L}}\langle 0|\mathcal{O}_{H}(z_{1})\mathcal{O}_{L}(z'_{2})(0)\mathcal{O}_{L}(z_{3})\mathcal{O}_{H}(z_{4})|0\rangle \times C.C 
			\end{equation}
			\normalsize
			In the above, we have used the fact that, in this case, the time evolution becomes a conformal transformation in the complex plane.
			\footnote{For details of the calculation see appendix \ref{appndxA} }. Our problem is now reduced to the computation of a four-point correlation function, which depends on the dynamical input of the CFT in consideration and is in general a complicated function $\mathcal{F}(\eta,\bar{\eta})$ of the cross-ratio $\eta$ and its complex conjugate $\bar{\eta}$ defined below,
			\begin{align*}
				\eta=\frac{z_{12}z_{34}}{z_{13}z_{24}}, \hskip 0.5cm \bar{\eta}=\frac{\bar{z}_{12}\bar{z}_{34}}{\bar{z}_{13}\bar{z}_{24}}  
			\end{align*}
			
			To compute $C_{\psi}$,  we again fix the initial positions of the operators on the cylinders as $\omega_{1}=T, \omega_{2}=0, \omega_{3}=0, \omega_{4}=-T$ and hence the positions on the plane are,
			\begin{align*}
				z_{1}=e^{T},\hskip 0.3cm z_{2}=1,\hskip 0.3cm z_{3}=1,\hskip 0.3cm z_{4}=e^{- T}
			\end{align*}
			In the $T \to \infty$ limit, the cross-ratio is
			\begin{align*}
				\eta=\frac{c_{\tau}+d_{\tau}}{a_{\tau}+b_{\tau}}, \hskip 0.5cm \bar{\eta}=\frac{\bar{c}_{\tau}+\bar{d}_{\tau}}{\bar{a}_{\tau}+\bar{b}_{\tau}}
			\end{align*}
			
			Putting the values of $z$, the auto-correlation function becomes,
			\small
			\begin{eqnarray}\label{ac}
				C_{\psi}(\tau)&=&\lim\limits_{T\to \infty} \frac{1}{\left|c_{\tau}+d_{\tau}\right|^{2 \Delta_{L}}}\left(e^{T}-e^{-T}\right)^{2 \Delta_{H}}\langle 0|\mathcal{O}_{H}(e^{T})\mathcal{O}_{L}\left(\frac{a_{\tau}+b_{\tau}}{c_{\tau}+d_{\tau}}\right)\mathcal{O}_{L}(1)\mathcal{O}_{H}(e^{-T})|0\rangle \times C.C \nonumber \\
				&=& \frac{1}{\left|c_{\tau}+d_{\tau}\right|^{2 \Delta_{L}}}\langle 0|\mathcal{O}_{H}(\infty)\mathcal{O}_{L}\left(\frac{1}{\eta}\right)\mathcal{O}_{L}(1)\mathcal{O}_{H}(0)|0\rangle \times C.C \nonumber\\
				&=& \frac{1}{\left|a_{\tau}+b_{\tau}\right|^{2 \Delta_{L}}}\mathcal{F}(\eta,\bar{\eta})
			\end{eqnarray}
			\normalsize
			In the second line of the above, we have used the fact that $\langle\mathcal{O}(\infty)|=\lim\limits_{\substack{z\to \infty \\ \bar{z}\to \infty}}\langle \mathcal{O}(z,\bar{z})|$ and we have redefined\footnote{We have used the identity $\langle\mathcal{O}_{H}(\infty)\mathcal{O}_{L}(\frac{1}{\eta}, \frac{1}{\Bar{\eta}})\mathcal{O}_{L}(1)\mathcal{O}_{H}(0)\rangle= |\eta|^{2 \Delta_{L}} \langle\mathcal{O}_{H}(\infty)\mathcal{O}_{L}(1)\mathcal{O}_{L}(\eta, \Bar{\eta})\mathcal{O}_{H}(0)\rangle $} $\eta$ as $\frac{1}{\eta}$. A closed-form analytic expression for $\mathcal{F}(\eta,\bar{\eta})$ is not known generally, except, in certain specific limits. One such example is when we consider that the central charge of the CFT is large ($c\to\infty$). We'll consider this scenario here.\\
			\vskip 0.1cm
			
			\subsubsection{For 2d CFT with large central charge}\label{extdLargeC}
			\vskip 0.1cm
   \begin{figure}
		\begin{subfigure}[h]{0.5\textwidth}
			\includegraphics[width=\textwidth]{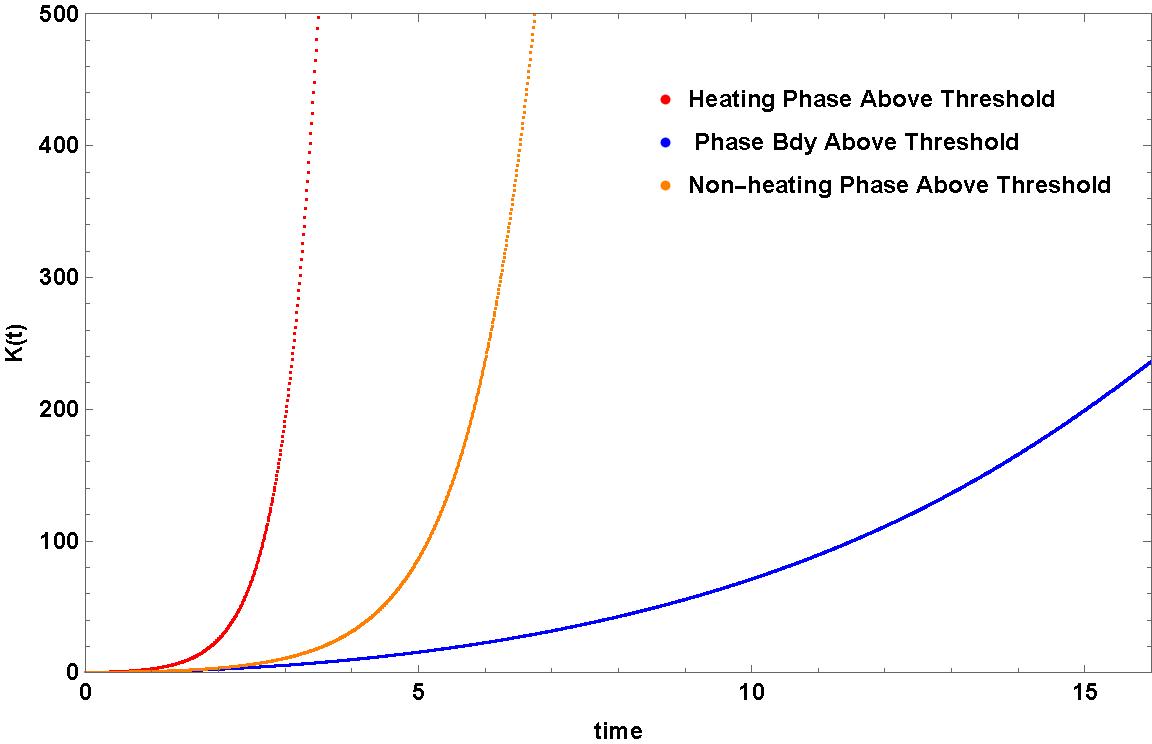}
			\caption{}
        \label{threephasesabove}
		\end{subfigure}
		\begin{subfigure}[h]{0.5\textwidth}
			\includegraphics[width=\textwidth]{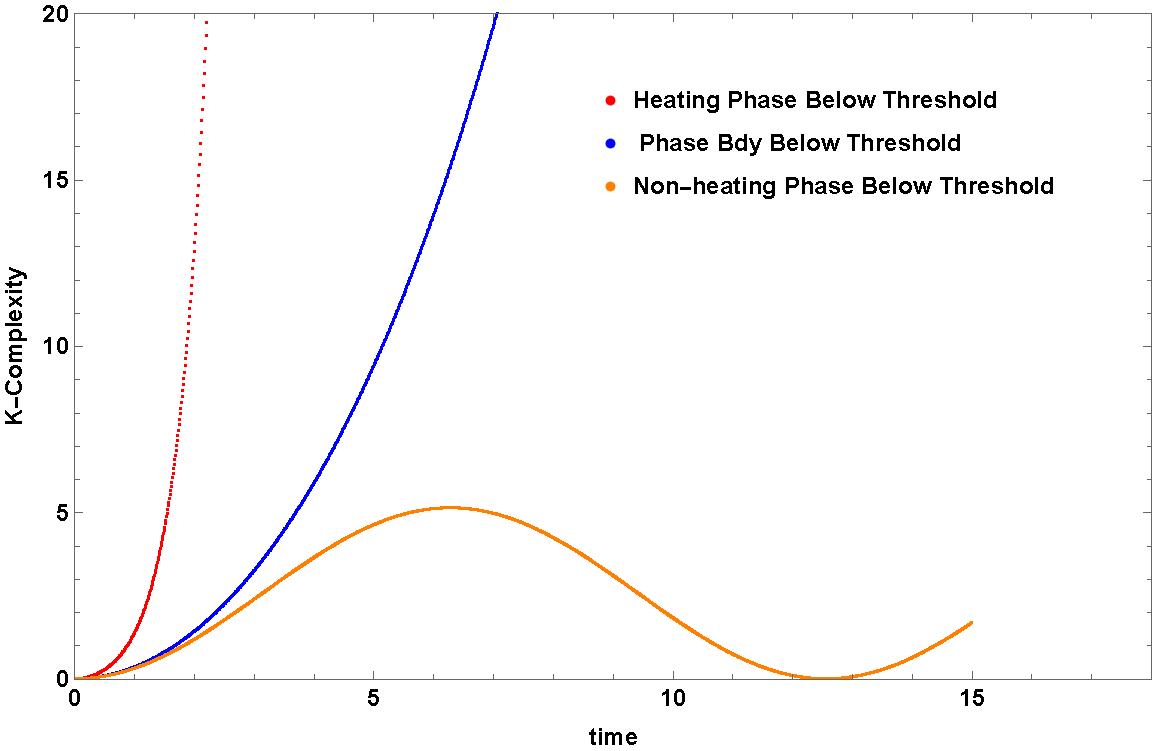}
			\caption{}
			\label{threephasesbelow}
		\end{subfigure}
		\caption{[a]. The k-complexity for Heating(\textcolor{red}{\textbullet}), Non-Heating(\textcolor{orange}{\textbullet}), and Phase-boundary(\textcolor{blue}{\textbullet}) when the weight of the heavy operator is above the black hole threshold. As is evident, in the Heating phase the complexity grows exponentially in time, while it remains polynomial for the Phase-boundary. However, in the Non-Heating phase, the behavior becomes exponential compared to the oscillatory behavior below the threshold. [b]. Below the BH threshold, the k-complexity for the heating phase(\textcolor{red}{\textbullet}), non-heating phase (\textcolor{orange}{\textbullet}), and phase boundary (\textcolor{blue}{\textbullet}) is exponential, oscillatory, and polynomial respectively similar to the vacuum state computation. }
		\label{fig:Adj-phaseVI-VII}
  \end{figure}
  
  \begin{figure}
      
		\begin{subfigure}[h]{0.5\textwidth}
			\centering
			\includegraphics[width=\textwidth]{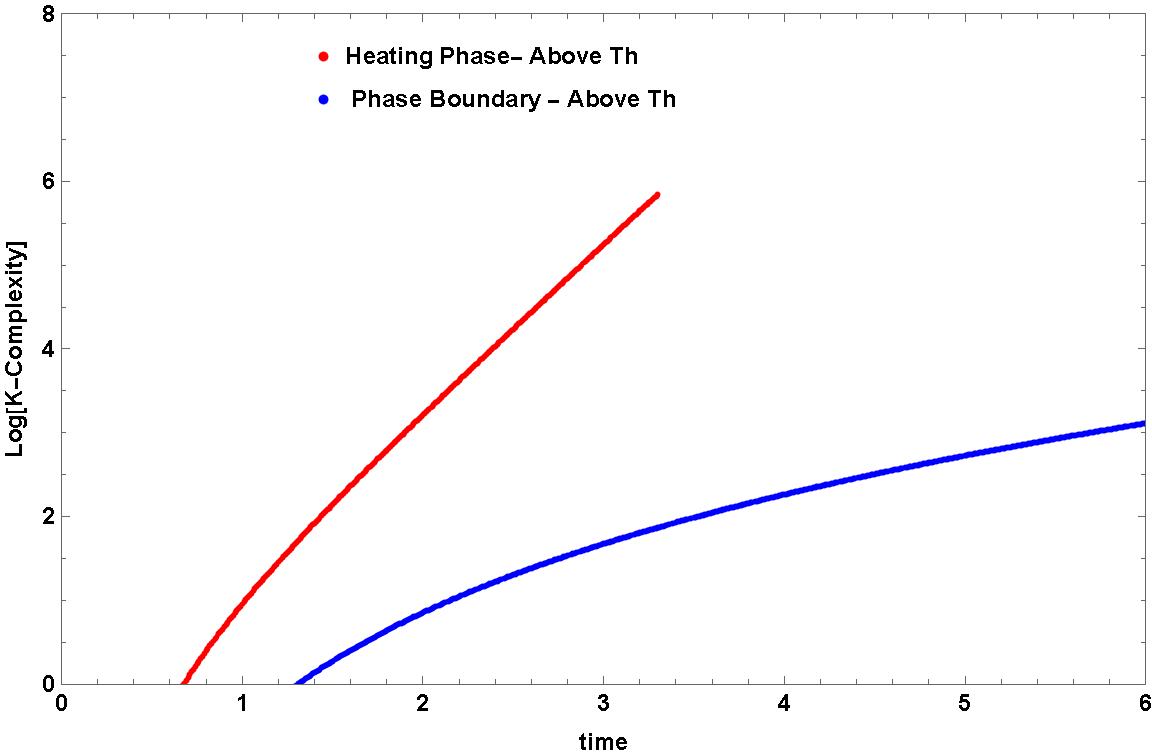}
			\caption{}
			\label{logabovetwo}
		\end{subfigure}
		\label{fig:Adj-phaseVI-VII}
		\begin{subfigure}[h]{0.5\textwidth}
			\centering
			\includegraphics[width=\textwidth]{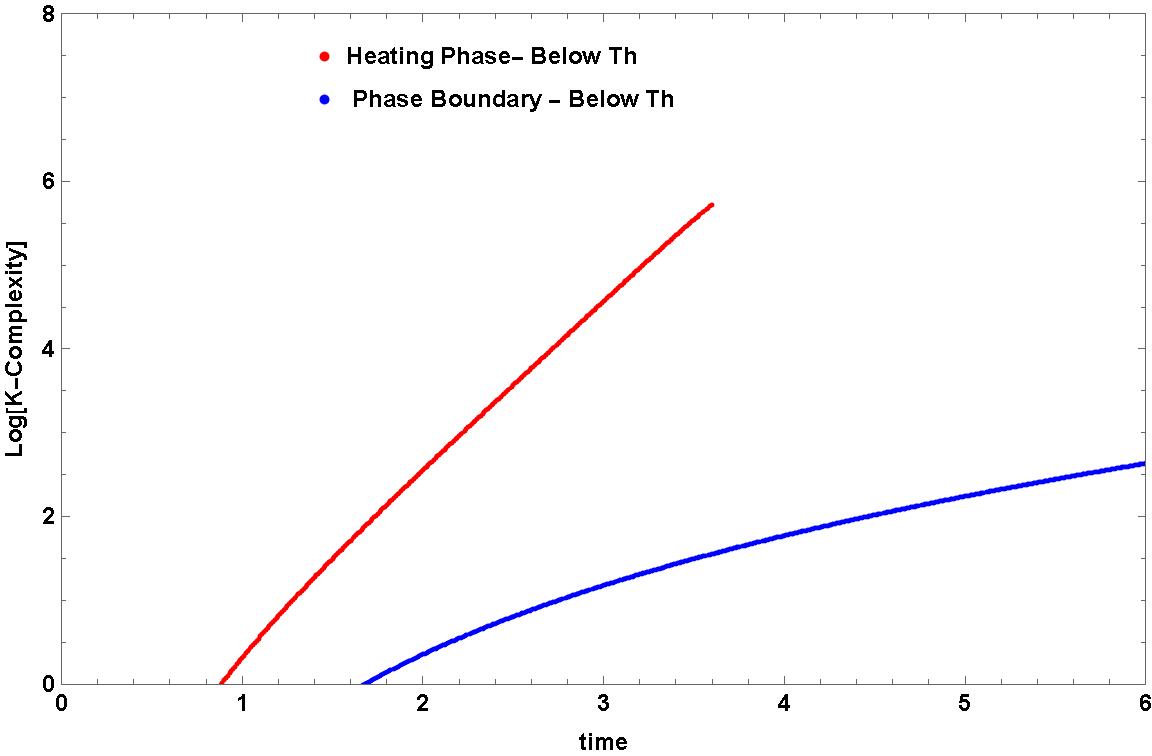}
			\caption{}
			\label{logbelowtwo}
		\end{subfigure}
		\caption{[a]. The logarithm of the complexity for the heating phase(\textcolor{red}{\textbullet}) and phase boundary(\textcolor{blue}{\textbullet}) above the threshold. As we can see, it saturates in the phase boundary but keeps increasing in the heating phase. [b].  The logarithm of the complexity for the heating phase(\textcolor{red}{\textbullet}) and phase boundary(\textcolor{blue}{\textbullet}) below the threshold. As we can see, it saturates in the phase boundary but keeps increasing in the heating phase, though the rate of increment decreases compared to the above threshold case.}
		\label{fig:Adj-phaseVI-VII}
	\end{figure}
\begin{figure}
      
		\begin{subfigure}[h]{0.5\textwidth}
			\includegraphics[width=\textwidth]{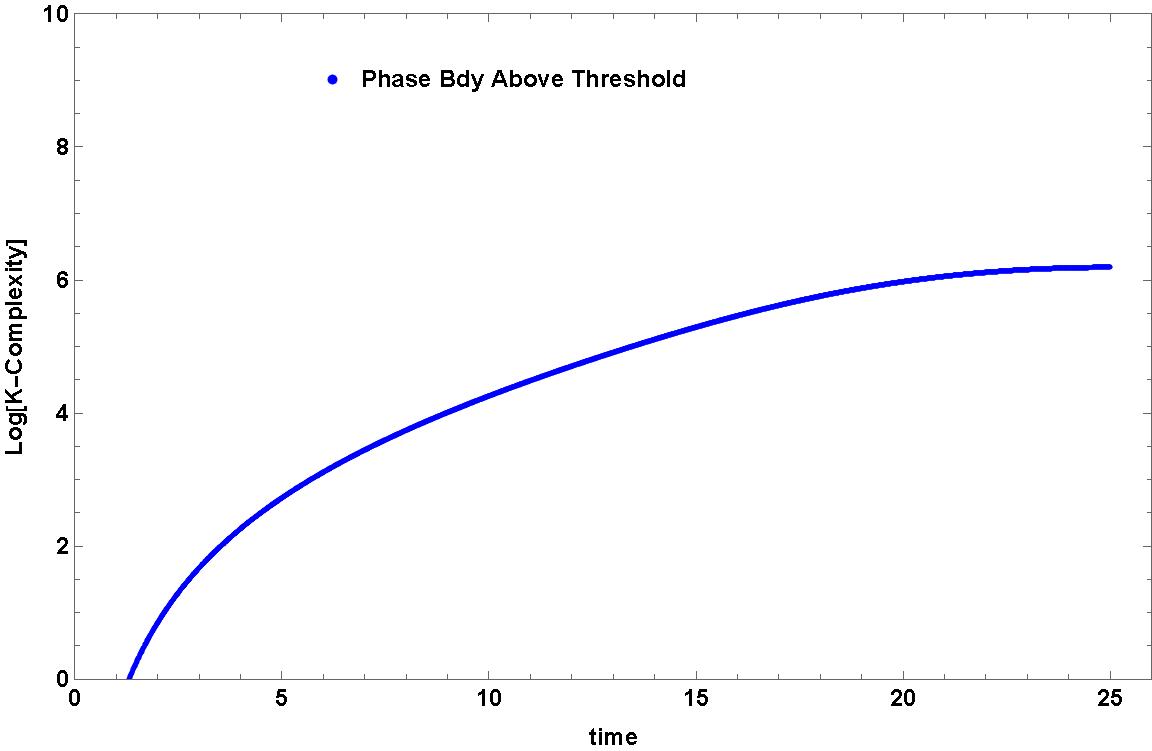}
			\caption{}
			\label{logpbabove}
		\end{subfigure}
		\label{fig:Adj-phaseVI-VII}
		\begin{subfigure}[h]{0.5\textwidth}
			\includegraphics[width=\textwidth]{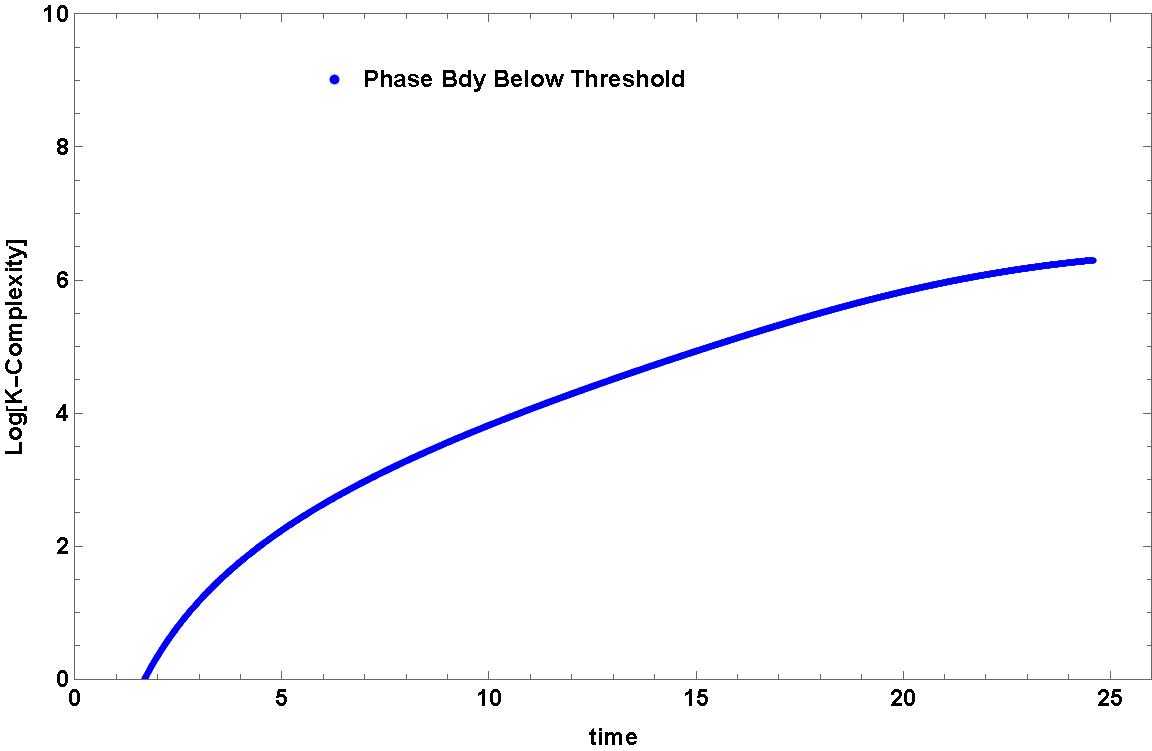}
			\caption{}
			\label{logpbbelow}
		\end{subfigure}
		\caption{[a]. The logarithm of the K-complexity in the phase boundary above the threshold. It saturates in a long time limit, confirming that the behavior is polynomial. [b].  The logarithm of the K-complexity in the phase boundary above the threshold. It saturates in a long time limit, which again confirms that the behavior is polynomial. }
		\label{fig:Adj-phaseVI-VII}
	\end{figure}

   It is known that any correlation function in CFT can be written as the sum of the Virasoro conformal blocks. However, when we consider $c$ to be large with $\frac{h_{L}}{c}$ is fixed and small and $h_{H}$ fixed and large, only the vacuum block dominates \cite{Fitzpatrick:2014vua,Fitzpatrick:2015zha}. In this limit, \vskip 0.1cm
			\begin{align}\label{4pt}
				\mathcal{F}(\eta,\bar{\eta})=\nu^{2 \Delta_{L}}\left(\frac{\eta^{\frac{\nu-1}{2}}}{1-\eta^{\nu}}\right)^{\Delta_{L}}\left(\frac{\bar{\eta}^{\frac{\nu-1}{2}}}{1-\bar{\eta}^{\nu}}\right)^{\Delta_{L}} +\mathcal{O}\left(\frac{1}{c}\right)\hskip 0.2cm \text{where} \hskip 0.2cm \nu=\sqrt{1-12\frac{\Delta_{H}}{c}}
			\end{align}
			Using (\ref{4pt}) in (\ref{ac}), the auto-correlation function becomes,
			\begin{align}
				C_{\psi}(\tau)= \left(\nu \right)^{2 \Delta_{L}}\left(\frac{(c_{\tau}+d_{\tau})^{\frac{\nu-1}{2}}(a_{\tau}+b_{\tau})^{\frac{\nu-1}{2}}}{(a_{\tau}+b_{\tau})^{\nu}-(c_{\tau}+d_{\tau})^{\nu}}\right)^{\Delta_{L}}\left(\frac{(\bar{c}_{\tau}+\bar{d}_{\tau})^{\frac{\nu-1}{2}}(\bar{a}_{\tau}+\bar{b}_{\tau})^{\frac{\nu-1}{2}}}{(\bar{a}_{\tau}+\bar{b}_{\tau})^{\nu}-(\bar{c}_{\tau}+\bar{d}_{\tau})^{\nu}}\right)^{\Delta_{L}}
			\end{align}
			The behavior of the auto-correlation function depends on $a_{\tau}, b_{\tau}, c_{\tau}, d_{\tau}$, and the parameter $\nu$. The parameters of the Möbius transformation depend on the phase we are considering, and details of the expressions are given in appendix \ref{appndxA}. In each phase, there are two different scenarios to consider separately: one when the weight of the heavy operator is above the critical value ($\Delta_{H}>\frac{c}{12}$) and the other case when it's below the critical value ($\Delta_{H}<\frac{c}{12}$).  
			In the previous case, we could write the analytical expressions of K-complexity in the three phases using the Lanczos algorithm and recursion relation in $ \phi_n$. For this case, however, it is difficult to find out the analytic expression for Lanczos coefficients and therefore of the K-complexity. One can however turn to numerical approaches to obtain some important observations in such cases.\\

			\textbf{I: Heating Phase ($\delta>0$)}\vskip 0.1cm
			
			We consider a particular choice of $\alpha, \beta$, and $\gamma$s that belong to the heating phase. We considered $\beta=1, \alpha=\gamma=0$. However, since all the other points within the heating phase are related to each other by some unitary transformation, the behavior would be similar for other choices as well.

With this particular choice, when the weight of the heavy operator is below the threshold \footnote{In this case and all the other cases, we have taken $\Delta_{L}=1$.} ($\Delta_{H}=\frac{3 c}{48}$), the auto-correlation function is,\\
        \begin{equation}
            C_{\psi}(\tau)= \frac{\left(\tan \left(\frac{\tau}{2}\right)+1\right)^{3/2} \cot \left(\frac{\tau}{2}\right)}{2 (\sin (\tau)+1) \sqrt{\tan \left(\frac{\tau}{2}\right)-1}}
        \end{equation}
         \begin{figure}[h!]
\centering
    \includegraphics[width=0.8\textwidth]{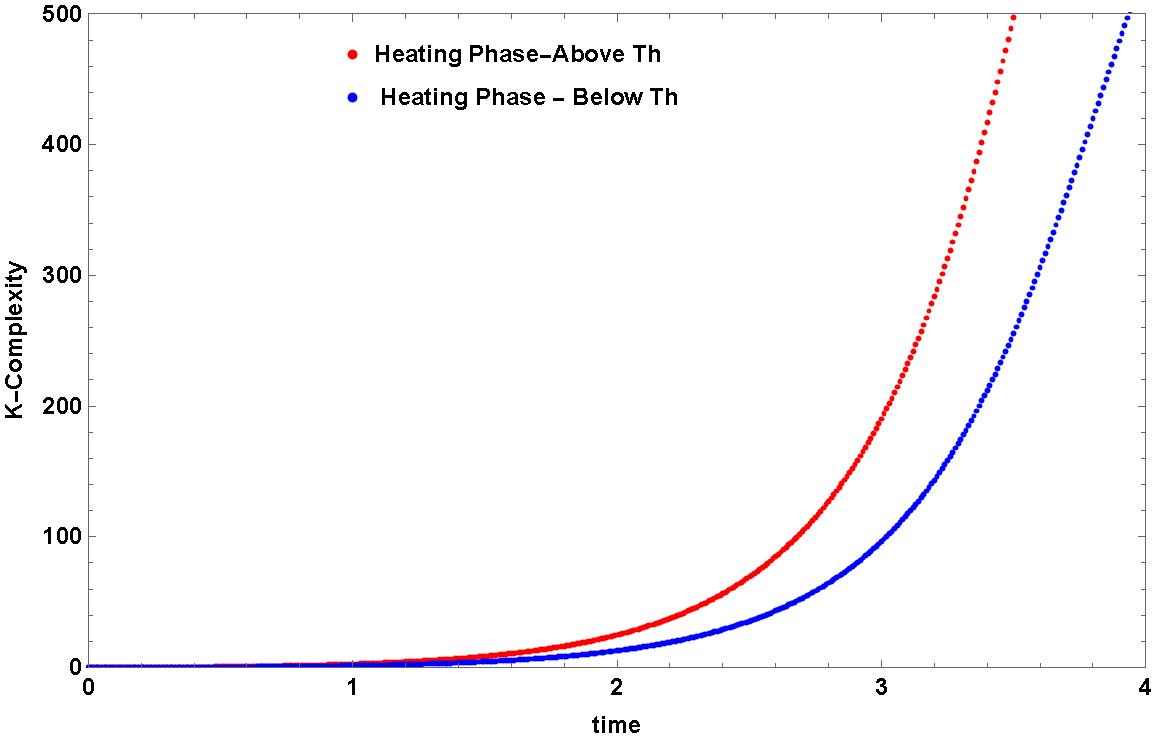}
    \caption{K-complexity in Heating phase above and below threshold}
    \label{fig:hpab}
\end{figure}
	With the above auto-correlation function, we find numerically that the Lanczos coefficients grow linearly in n. The Krylov complexity grows exponentially in time, as shown in Fig.[\ref{fig:hpab}]. On the other hand, when the weight of the heavy operator is above the critical value ($\Delta_{H}=\frac{c}{6}$), the correlation function is\\
 \begin{equation}
     C_{\psi}(\tau)=\frac{\sin (\tau )+\cos (\tau )+1}{2 (\sin (\tau )+1) (\sin (\tau )-\cos (\tau )-1) \left(\cos \left(\frac{1}{2} \log \left(\frac{2}{\sin (\tau )+1}-1\right)\right)-\cosh (\pi )\right)}
 \end{equation}
In this case also the behavior of the Lanczos coefficients and the Krylov complexity is similar to the above case. However, we notice that the rate of the growth of the complexity increases in this scenario (Fig.[\ref{fig:hpab}]). To confirm that the behavior of the K-complexity is indeed exponential both above and below the threshold, we also plot the logarithm of the K-complexity in both cases, and we find that it keeps increasing linearly in time (Fig.[\ref{logabovetwo}, \ref{logbelowtwo}]. \\

\textbf{II:  Non-heating phase($\delta<0$)}\vskip 0.1cm
The simplest example of the non-heating phase is when $\alpha=1$ and $\beta=\gamma=0$. This choice leads to the usual CFT Hamiltonian. In this case, the expressions of the auto-correlation functions above and below-threshold are much simpler and hence we can get the exact analytic expressions of Lanczos coefficients and the complexity \cite{Kundu:2023hbk}.
\begin{figure}[h!]
\centering
    \includegraphics[width=0.8\textwidth]{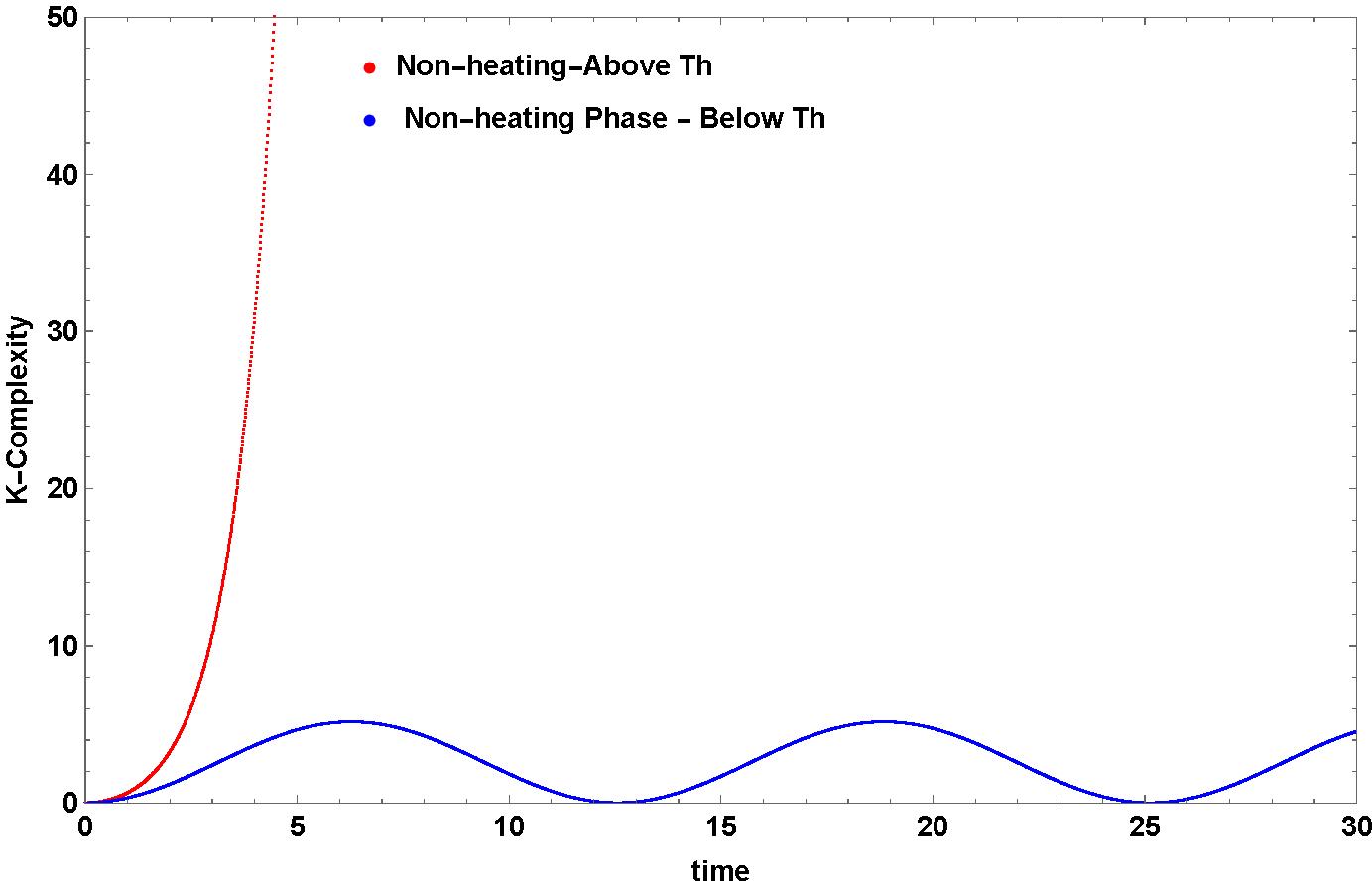}
    \caption{K-complexity in non-heating phase above and below threshold}
    \label{fig:pbab}
\end{figure}
When the weight of the heavy operator is below the threshold, the auto-correlation function is\\
\begin{equation}
    C_{\psi}(\tau)=\frac{(\frac{\gamma}{2})^{2 \Delta_{L}}}{\sinh{\left(\frac{\gamma \tau}{2}\right)}^{2 \Delta_{L}}}
\end{equation}
In this case, the k-complexity is given by
\begin{equation}
    K_{L}(t)=2 \Delta_{L} csch^{2}(\frac{\gamma \tau_0}{2})\sin^2{\left(\frac{\gamma t}{2}\right)}
\end{equation}
While above the threshold, the auto-correlation 
function is,
\begin{equation}
    C_{\psi}(\tau)=\frac{(\frac{\gamma}{2})^{2 \Delta_{L}}}{\sin{\left(\frac{\gamma \tau}{2}\right)}^{2 \Delta_{L}}}
\end{equation}
With the above auto-correlation function, the complexity is,
\begin{equation}
    K_{L}(t)=2 \Delta_{L} csc^{2}(\frac{\gamma \tau_0}{2})\sinh^2{\left(\frac{\gamma t}{2}\right)}
\end{equation}
So, as one can see, there is a change in the behavior of the complexity from oscillatory to exponential when we go from below to above the threshold limit. As discussed in \cite{Kundu:2023hbk}, this change in the behavior is related to the fact that two phases correspond to thermal AdS$_{3}$ and black hole geometries in the bulk. \\ 

\textbf{III: Phase boundary ($\delta=0$)}\vskip 0.1cm
To study the behavior in the phase boundary, we choose $\alpha=\beta=1$ and $\gamma=0$.
  \begin{figure}[h!]
\centering
    \includegraphics[width=0.8\textwidth]{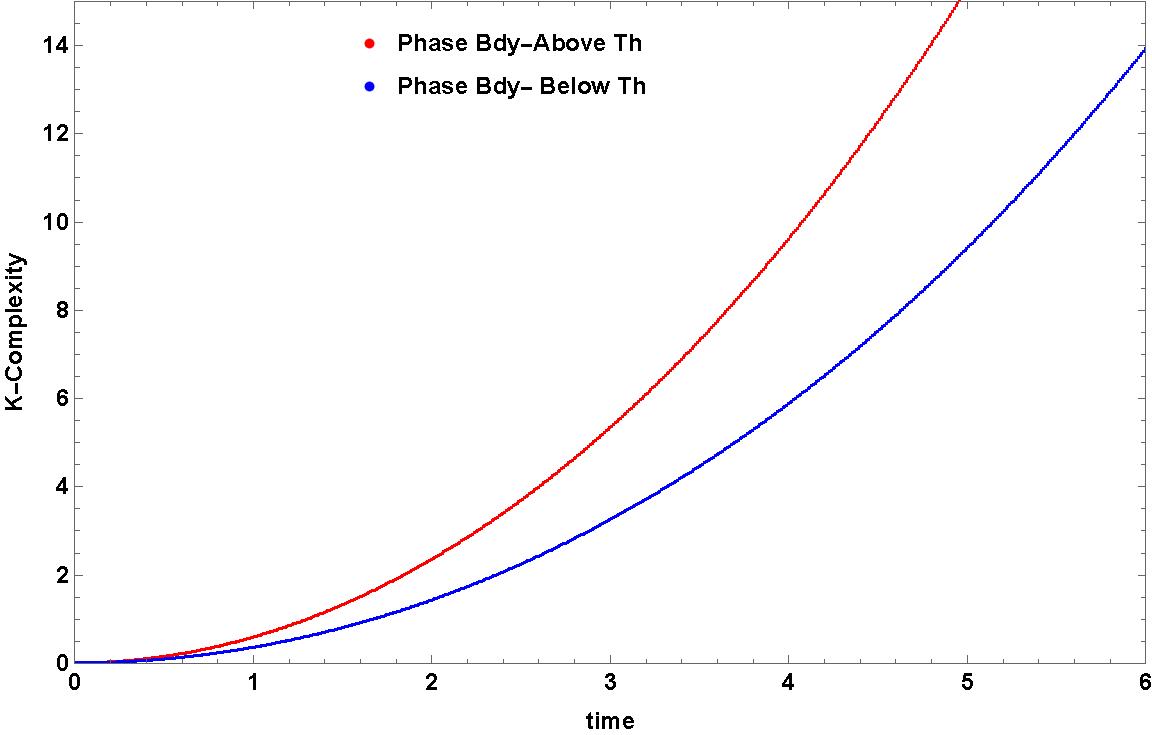}
    \caption{K-complexity in phase boundary above and below threshold}
    \label{fig:pbab}
\end{figure}
With this choice, we find that in the case when we are below threshold, the auto-correlation function is:
\begin{equation}
     C_{\psi}(\tau)=\frac{1}{2 \tau \sqrt{\tau^2-1}}
\end{equation}
With the above expression, we find that again the Lanczos coefficients grow linearly in n as expected, but this time the complexity grows polynomial in time. The auto-correlation function above the threshold value is, 
\begin{equation}
    C_{\psi}(\tau)=-\frac{\cos \left(\log \left(\frac{\tau+1}{\tau-1}\right)\right)+\cos \left(\log \left(\frac{\tau-1}{\tau+1}\right)\right)-2 \cosh (\pi )}{2 \left(\tau^2-1\right) \left((4 \cosh (\pi )-1) \cos \left(\log \left(\frac{\tau-1}{\tau+1}\right)\right)-2-\cosh (2 \pi )\right)}
\end{equation}
In this scenario, we once again notice a polynomial increase in K-complexity. However, it is noteworthy that the growth rate intensifies during this phase (Fig.[\ref{fig:pbab}]). To confirm that the behavior in both cases is polynomial, we also plot the logarithm of the K-complexity in both cases, and we observe that it saturates to a particular value after a certain time (Fig.[\ref{logpbabove}], Fig.[\ref{logpbbelow}]). Due to limitations in numerical computations, we couldn't plot it for longer times, but the plots clearly indicate that the behavior is polynomial.  \\
To compare the behavior of the K-complexity in all three phases, both above and below the threshold, we have plotted them together in Fig.[\ref{threephasesabove}] and Fig.[\ref{threephasesbelow}].

\subsubsection{For massless free scalar field theory}\label{extdscalar}
\vskip 0.1cm
Once more, we examine the auto-correlation function in the excited state. This time, our focus is on the two-point function of two scalar operators with $\Delta_{1}=1$. The state is generated by applying the vertex operator $e^{i k \phi}$ to the vacuum. With this specific operator configuration, the auto-correlation function in the deformed CFT case is derived as follows:
\begin{equation}{\label{autofree}}
   C_{\psi}(\tau)=\lim\limits_{T\to \infty}\frac{1}{\left|c_{\tau} z_{2}+d_{\tau}\right|^{2 \Delta_{L}}}(z_2 z_3)\left(\frac{1}{z_{12\tau}z_{34}}+\frac{1}{z_{13}z_{2\tau4}}-\frac{1}{4\pi}\frac{1}{z_{2\tau3}^2}\right)
\end{equation}
Where, $z_{ij}=z_i-z_j$ and $z_{i\tau}=\frac{a_{\tau}z_i+b_{\tau}}{c_{\tau}z_i+d_{\tau}}$.
The exact form of \eqref{autofree} depends on the form of $a_{\tau}, b_{\tau}, c_{\tau}$ and $d_{\tau}$ and hence on the phase in which we are computing. For the sake of simplicity, we consider specific choices of $\alpha, \beta$, and $\gamma$s as representatives of different phases. For example, we choose $\alpha \neq 0, \beta=\gamma=0$ for non-heating phase and $\beta \neq 0, \alpha=\gamma=0$ and $\alpha=\beta \neq 0, \gamma=0$ for heating phase and phase boundary respectively.\\
\textbf{Non-heating phase:}
With the above choices of $\alpha, \beta, \gamma$, the auto-correlation function is,\\
\begin{equation}
    C_{\psi}(\tau)=-\frac{1}{16 \pi}\left(\frac{1}{\sinh{\frac{\alpha \tau}{2}}}\right)^2
\end{equation}
\textbf{Heating phase:}
In the heating phase, the auto-correlation function is
\begin{equation}
    C_{\psi}(\tau)=-\frac{1}{16 \pi}\left(\frac{1}{\sin{\frac{\beta \tau}{2}}}\right)^2
\end{equation}
\textbf{Phase boundary:}
In the phase boundary, the auto-correlation function is
\begin{equation}
    C_{\psi}(\tau)=-\frac{1}{16 \pi}\left(\frac{1}{\beta \tau}\right)^2
\end{equation}
As we can see from the above expression, the auto-correlation function matches with the auto-correlation functions in the vacuum state (\eqref{twopthp}-\eqref{twoptpb}), with $\Delta=1$.
Therefore, the Lanczos coefficients and the Krylov complexities will be similar to those in the three phases. The important point to notice here is that, even in the free field theory case, Krylov complexity grows exponentially in time in the heating phase. Hence, the K-complexity can't differentiate between a chaotic (large-c) and non-chaotic CFT (free field theory) in the heating phase and on the phase boundary.	
		
		\numberwithin{equation}{section}

  \section{Discussion}\label{dissc}





In this article, we studied the evolution of operator complexity within the framework of $SL(2,\mathbb{R})$ deformed Conformal Field Theories (CFTs) in 1+1 dimensions. This investigation is equally applicable to the study of complexity in the context of periodically driven/floquet CFTs, given that the dynamics of such systems can be effectively described by the Hamiltonian of these deformed CFTs. Our findings reveal distinct qualitative behaviors of K-complexity in different phases.

During the non-heating phase, K-complexity demonstrates oscillatory behavior. In contrast, within the phase boundary and heating phases, particularly when considering the vacuum state of the CFT, we observe both polynomial and exponential growth. However, when examining a generic state in the non-heating phase of large c CFT, K-complexity behaves differently above and below the threshold. Above the threshold, complexity exhibits exponential growth, while below the threshold, it displays oscillatory behavior. In the non-heating phase of a free-field theory, the dynamics of K-complexity always remain oscillatory.

Interestingly, it appears that in some generic states during the non-heating phase, K-complexity can differentiate between chaotic and non-chaotic CFTs. Nevertheless, in the heating phase and at the phase boundary, the situation changes. Notably, in large c, the behavior of complexity remains exponential and polynomial both above and below the threshold in these phases. Even in the massless free field theory, complexity displays exponential and polynomial growth in the heating phase and at the phase boundary.

The inability of K-complexity to differentiate between chaotic and non-chaotic CFTs in the heating phase and at the phase boundary may be attributed to the fact that distinct phases correspond to different quantizations of 2D CFT. For instance, the sine-squared deformed Hamiltonian, which is an example of phase boundary, corresponds to `dipolar quantization,' where the spectrum of Hamiltonian is continuous due to the infinite effective length of the system \cite{Ishibashi:2015jba}. A similar continuous spectrum is observed in the heating phase as well \cite{Tada:2019rls}. This contrasts with radial quantization, which yields a discrete spectrum. Thus, K-complexity seems to differentiate between chaotic and non-chaotic systems only when the theory exhibits a discrete spectrum, aligning with the observations in \cite{Iizuka:2023pov}.

As a future direction, it would be interesting to extend this study to other quench protocols as well. For instance, one immediate extension could be to analyze a global quench protocol where one initially starts from a regulated boundary state and then evolves the system with the usual CFT Hamiltonian. It was discussed in \cite{cardy}, that such a system thermalizes at late times. Therefore, it would be interesting to study how the K-complexity evolves in this system.
In \cite{Rabinovici:2023yex}, the authors conjectured a holographic bulk dual description of K-complexity in terms of the length of a two-sided wormhole. Since, the holographic dual geometries of $SL(2,\mathbb{R})$ deformed CFTs are known \cite{Das:2022pez}, it would be worthwhile to study the complexity from the bulk and match it with the boundary results.

   \vskip 0.8cm
\section*{Acknowledgments}

The authors thank Suman Das for the initial collaboration on some topics related to this project. The authors are thankful to Bobby Ezhuthachan, Arnab Kundu, Ritam Sinha, and Junggi Yoon for several useful discussions and comments. 
 BR would like to thank YITP, Kyoto University for their warm hospitality at the long-term conference (YITP-T-23-01), during which a part of this work was done. SP would like to thank Prof. Junggi Yoon for the month-long visit to APCTP, Pohang, where, some final works of this project were completed.
The work of VM was supported by the NRF grant funded by the Korea government (MSIT) (No. 2022R1A2C1003182) and by the Brain Pool program funded by the Ministry of Science and ICT through the National Research Foundation of Korea (RS-2023-00261799). The work of SP and BR is supported by the Senior Research Fellowship(SRF) funded by the University Grant Commission(UGC) under the CSIR-UGC NET Fellowship.
   
   \appendix
   
   \section{ Conformal transformation generated by the deformed Hamiltonian}\label{appndxA}

In this section, we briefly review the dynamics of a deformed conformal field theory. Consider a 1+1D CFT defined on an infinite cylinder of circumference L($=2 \pi$). The Hamiltonian of the system generates translations along the length of the cylinder and therefore is defined by integrating the energy density on a time slice $t_c=0$ over the spatial circle $x_c=x$\footnote{The subscript c denotes the coordinates on the cylinder.} as

\begin{equation}
    H=\oint T_{00}(x) dx
\end{equation}

Using analytic continuation to Euclidean space on a complex plane defined by $\zeta=\tau_c + i x_c$ and $\bar{\zeta}=  \tau_c- i x_c$ ) the above equation can be rewritten in terms of the left(right) moving components $T(\bar{T})$ of the energy-momentum tensor and the corresponding Virasoro modes 

\begin{align}\label{radial}
     H &=\frac{1}{2 \pi}\oint (T(\zeta) d\zeta + \bar{T} (\bar{\zeta})d\bar{\zeta})\\
     &=  L_0 + \bar{L}_0 - \frac{c}{12}\nonumber
\end{align}

using the usual constructions of operator formalism in radial quantization.



In the case of a Deformed CFT, instead of evolving the system with an ordinary CFT Hamiltonian H, one studies the operator evolution under a deformed CFT Hamiltonian 
\begin{equation}
    H_{def} = \oint f(x) T_{00} (x) + g(x) T_{01} (x)  dx
\end{equation}

where the energy density $T_{00}$ and momentum density $T_{01}$ are changed using some control function $f(x)$ and $g(x)$. For simplicity, one can make the choice $f(x)=g(x)$ and therefore
\begin{equation}
    H_{def} = \frac{1}{2 \pi}\oint f(\zeta) T(\zeta) d\zeta + \bar{f}(\bar{\zeta})\bar{T} (\bar{\zeta})d\bar{\zeta}
\end{equation}.

This deformed Hamiltonian $H_{def}$ in general can be written as a linear combination of Virasoro generators $L_n$s and $\bar{L}_n$s. 
\begin{equation*}
         H=\sum_n p_n L_n+\sum_n q_n\overline{L}_n- \frac{c}{12}
    \end{equation*}
One can simplify the problem by restricting to a certain choice of functions such that the deformed Hamiltonian is only made out of the generators SL$_q(2,\mathcal{R})$,i.e. $(L_0, L_q, L_{-q})$ and its conjugates, as described in \eqref{defh}
 
\begin{align}\label{defh3}
				H_{def}=\alpha(L_{0}+\bar{L}_{0})+\beta (L_{q+}+\bar{L}_{q+})+ \gamma (L_{q-}+\bar{L}_{q-})- \frac{c}{12}
			\end{align}

where, $L_{q+}=\frac{L_{q}+L_{-q}}{2}$, $L_{q-}=\frac{L_{q}-L_{-q}}{2 i}$ and $\alpha,\ \beta $ and $\gamma$ are some real numbers.
This linear combination of generators can be classified into three distinct classes, depending on the quantity $\delta=-\alpha^2+\beta^2+\gamma^2$ defined using the coefficients $\alpha,\ \beta,\ \gamma$. This can be understood as the adjoint action on $\alpha L_{0}+\beta  L_{q+} + \gamma L_{q-}$ changes the coefficients to some other values, $\alpha',\ \beta',\ \gamma'$ keeping the value of $\delta$ same.
\begin{equation*}
    \delta=-\alpha^2+\beta^2+\gamma^2=-\alpha'^2+\beta'^2+\gamma'^2
\end{equation*}
The three classes of Hamiltonians for $\delta\  (=0,\ >0,\ <0)$ are known respectively as parabolic, hyperbolic, and elliptic in the literature. It has been shown that two-point and one-point correlation functions e.g. entanglement entropy and energy density show different dynamical behaviour when evolved with these three types of Hamiltonians. The EE shows linear, oscillatory, and logarithmic behavior whereas the Energy density shows exponential, oscillatory, and linear behaviour for $\delta>0,<0$, and $\delta=0$ respectively. Following the authors in \cite{Wen:2018agb,Wen:2020wee} we will be referring to the three classes as heating $(\delta>0)$ phase and non-heating phase $(\delta<0)$ separated by a phase boundary $(\delta=0)$.

In \cite{Das:2022jrr} it has also been shown that even non-trivial higher point correlators such as four-point OTOC show different behaviors in these three phases.  

The fact that we can now write the Hamiltonian in $\eqref{defh3}$ as a linear combination of Virasoro generators, the dynamics get mapped to a conformal transformation and therefore the problem becomes analytically tractable. Specifically, in the Heisenberg picture, the time evolution (with the Euclidean time $s$\footnote{Note that, here we refer to this Euclidean time "s" (instead of $\tau$ )}) of a primary operator (with conformal weights {$\mathbf{h}$, $\mathbf{\overline{h}}$}), gets translated into the dynamics of the operator under conformal transformation
\begin{eqnarray}
e^{ H_{def} s} O(z, {\overline z}) e^{- H_{def} s} &=& \left( \frac{\partial
z'}{\partial z}\right)^h \left( \frac{\partial {\overline z'}}{\partial
{\overline z}}\right)^{\overline h} O(z',{\overline z'}) \label{opevoldef}
\end{eqnarray}

in that maps the coordinates $z=e^{\frac{2 \pi q \zeta}{L} },\bar{z}=e^{\frac{2 \pi q \bar{\zeta}}{L} }$ to some $(z',\bar{z}')$ by generating M\"{o}bius transformation in the q-sheet Riemann surface\footnote{ The Conformal map  $z=e^{\frac{2 \pi q \zeta}{L} },\bar{z}=e^{\frac{2 \pi q \bar{\zeta}}{L} }$ maps the cylinder coordinates $(\zeta,\bar{\zeta})$ to the coordinates $(z,\bar {z})$ of the q-sheet surface defined by $z$. For q=1, this  corresponds to the usual map from cylinder to the complex plane $(z,\bar{z})$}\footnote{we have only written the holomorphic part}

\begin{eqnarray}
z \to z'= \frac{a_s z+b_s}{c_s z+d_s} \ \ \text{with} \ \ \left(\begin{array}{cc}  a_s & b_s\\ c_s & d_s \end{array} \right) \label{ueq2} \in {SL}(2, {\mathbb R}) 
\end{eqnarray}

This enables us to now use the usual rules of 2D CFT at our disposal to find out how correlation functions behave under such conformal transformations. Our next task, therefore would be to find out the analytic expressions for $a_s, b_s, c_s, d_s$ in terms of the Hamiltonian parameters $\alpha,\beta,\gamma,\delta$ and Euclidean time $s$.

An easy method to do so is to re-write the Hamiltonian as a generator of translation in a coordinate system labeled by $(\tilde{z},\bar{\tilde{z}})$, such that,

\begin{eqnarray}
    H_{def}= \tilde{\mathcal{L}}_{-1} + \bar{\tilde{\mathcal{L}}}_{-1}\ \  \  \  \text{and}\ \ \ \ \ e^{s H_{def}} \tilde{z}\implies \tilde{z}'= \tilde{z}+s
\end{eqnarray}
Using this and by expressing the coordinate transformations written in terms of $(\tilde{z},\bar{\tilde{z}})$ and $(\tilde{z}',\bar{\tilde{z}'})$ to write in terms of original coordinates $(z,\bar{z})$ and $(z',\bar{z'})$, one can find the expressions for $a_s,b_s,c_s,d_s$ to be as follows:\\

For heating Phase
\begin{align}\label{HPab}
    a_s=-\frac{\alpha}{\sqrt{|\delta|} }  \sin \left(\frac{\sqrt{|\delta|} q s }{2}\right)-\cos \left(\frac{\sqrt{|\delta|} q s }{2}\right),\ \ \ \ \ \ b_s=-\frac{(\beta +i \gamma ) \sin \left(\frac{\sqrt{|\delta|} q  s }{2}\right)}{\sqrt{|\delta|} }\\
    c_s= \frac{(\beta -i \gamma ) \sin \left(\frac{\sqrt{|\delta|} q  s }{2}\right)}{\sqrt{|\delta|} },\ \ \ \  \ \ \ d_s=-\frac{\alpha}{\sqrt{|\delta|} }  \sin \left(\frac{\sqrt{|\delta|} q s }{2}\right)+\cos \left(\frac{\sqrt{|\delta|} q s }{2}\right)\nonumber
\end{align}\vskip 0.5cm

For non-heating Phase
\begin{align}\label{NHPab}
    a_s=-\frac{\alpha}{\sqrt{|\delta|} }  \sinh \left(\frac{\sqrt{|\delta|} q s }{2}\right)-\cosh \left(\frac{\sqrt{|\delta|} q s }{2}\right),\ \ \ \ \ \ b_s=-\frac{(\beta +i \gamma ) \sinh \left(\frac{\sqrt{|\delta|} q  s }{2}\right)}{\sqrt{|\delta|} }\\
    c_s= \frac{(\beta -i \gamma ) \sinh \left(\frac{\sqrt{|\delta|} q  s }{2}\right)}{\sqrt{|\delta|} },\ \ \ \ \ d_s=-\frac{\alpha}{\sqrt{|\delta|} }  \sinh \left(\frac{\sqrt{|\delta|} q s }{2}\right)+\cosh \left(\frac{\sqrt{|\delta|} q s }{2}\right)\nonumber
\end{align}
\vskip 0.5cm 
For Phase boundary
\begin{align}\label{PBab}
    a_s= -(1+ \frac{\alpha \sqrt{|\delta|} q s}{2})\ \ \ \ \ \ b_s=- \frac{(\beta +i \gamma ) \sqrt{|\delta|} q s}{2})\\ 
    c_s= \frac{(\beta -i \gamma ) \sqrt{|\delta|} q s}{2})\ \ \ \ \ \ \ d_s=(\frac{\alpha \sqrt{|\delta|} q s}{2}-1)\nonumber 
\end{align}
where for each case $a_s d_s- b_s c_s=1$ is satisfied.
In this paper, we have simplified our problem further by considering the q=1 case, such that, the Hamiltonian is built out of only the global Virasoro generators $L_0,L_{\pm1}$ and their conjugates. One must note that these generators keep the vacuum state invariant, and therefore for the initial vacuum state only unequal time correlators can capture the dynamics of the drive.

\section{Numerical Technique utilized to compute the K-Complexity}\label{appndxB}

\begin{figure}
		\begin{subfigure}[h]{0.5\textwidth}
			\includegraphics[width=\textwidth]{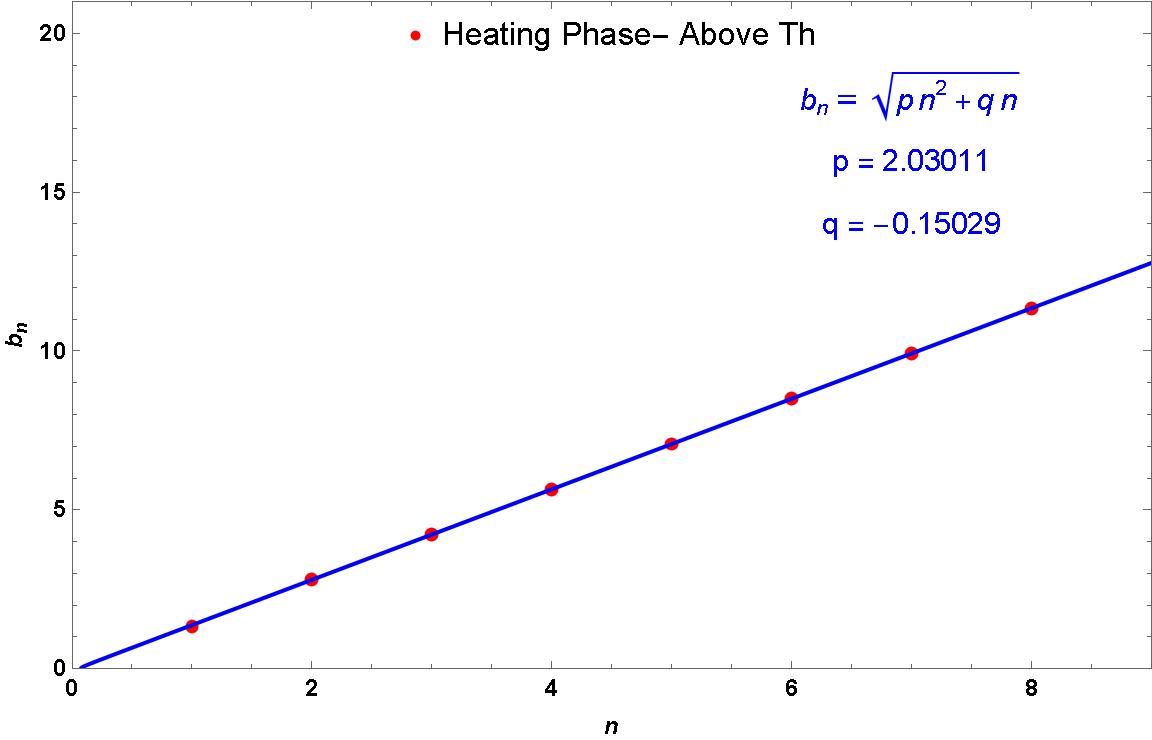}
			\caption{}
        \label{alanczosheatingphasesabove}
		\end{subfigure}
		\begin{subfigure}[h]{0.5\textwidth}
			\includegraphics[width=\textwidth]{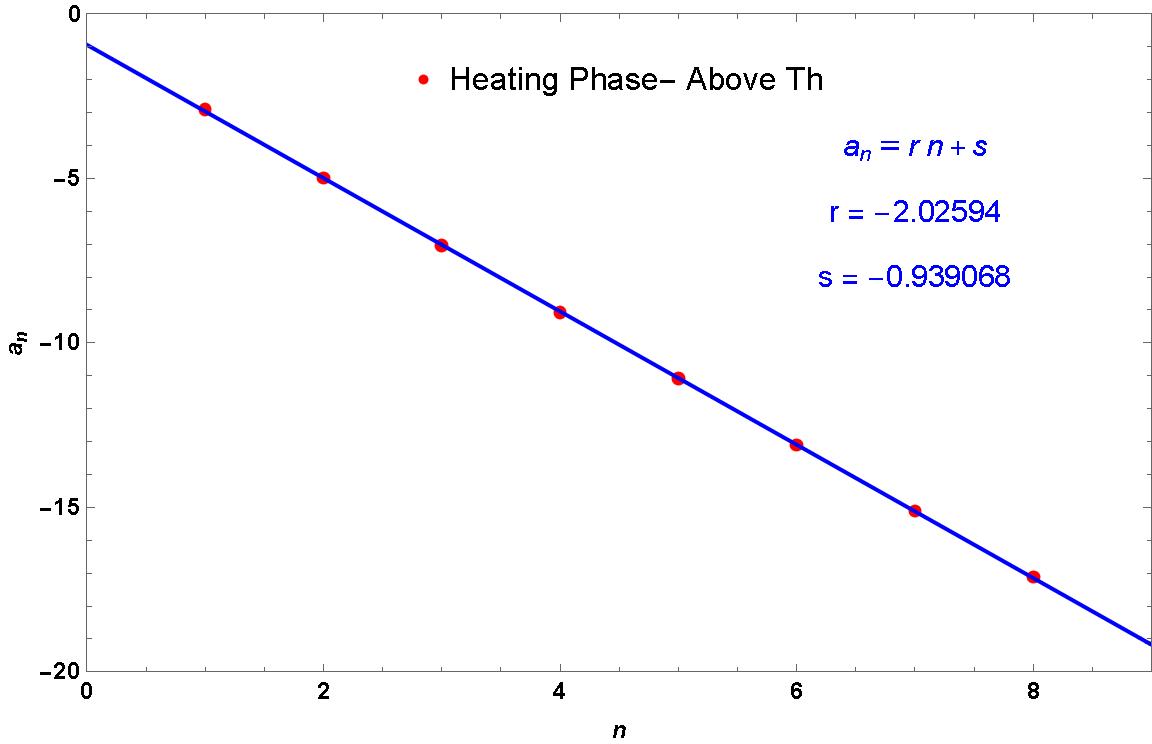}
			\caption{}
			\label{blanczosheatingphasesabove}
		\end{subfigure}
		\caption{The above plots show the behavior of the Lanczos coefficients in the Heating phase above the threshold value. Plot [a]. shows Lanczos coefficient $b_n$ increases linearly in $n$ and [b]. shows Lanczos coefficient $a_n$ decreases with $n$. The red dots are the numerical data points, and the blue line represents the best fit, determined by specific values of the parameter sets (p, q) and (r, s).}
		\label{lanczosheating} 
  \end{figure}

\begin{figure}
		\begin{subfigure}[h]{0.5\textwidth}
			\includegraphics[width=\textwidth]{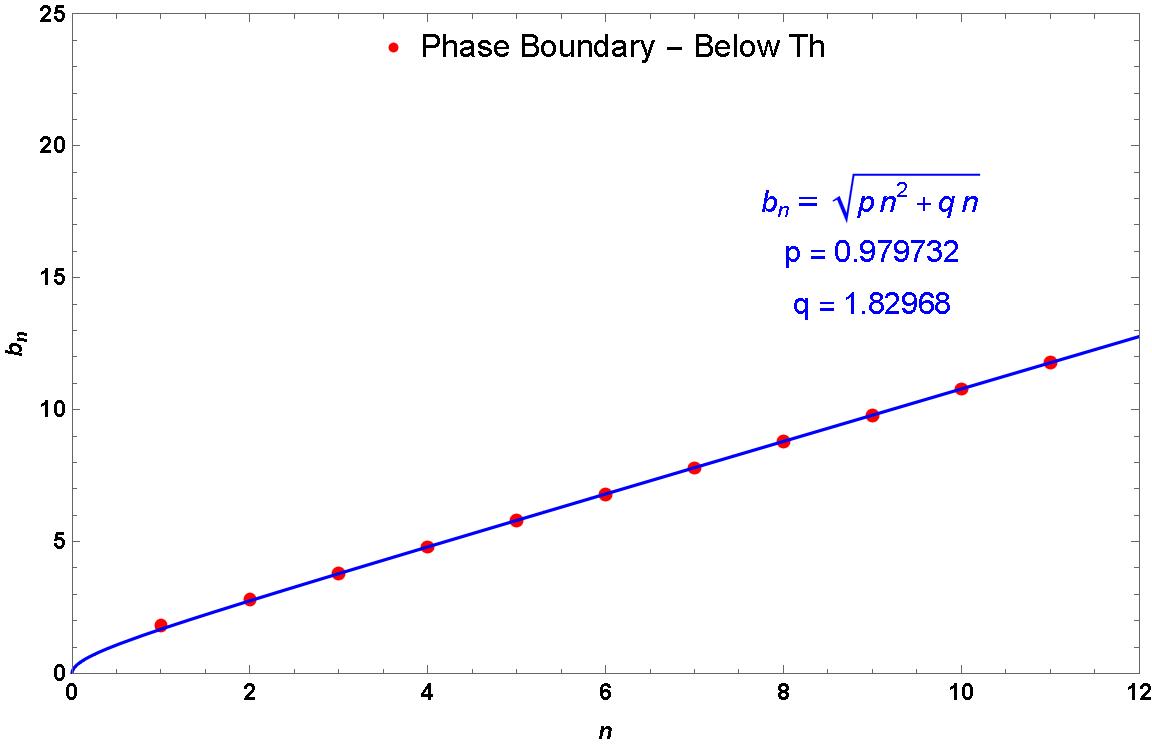}
			\caption{}
        \label{alanczosphasesbdbelow}
		\end{subfigure}
		\begin{subfigure}[h]{0.5\textwidth}
			\includegraphics[width=\textwidth]{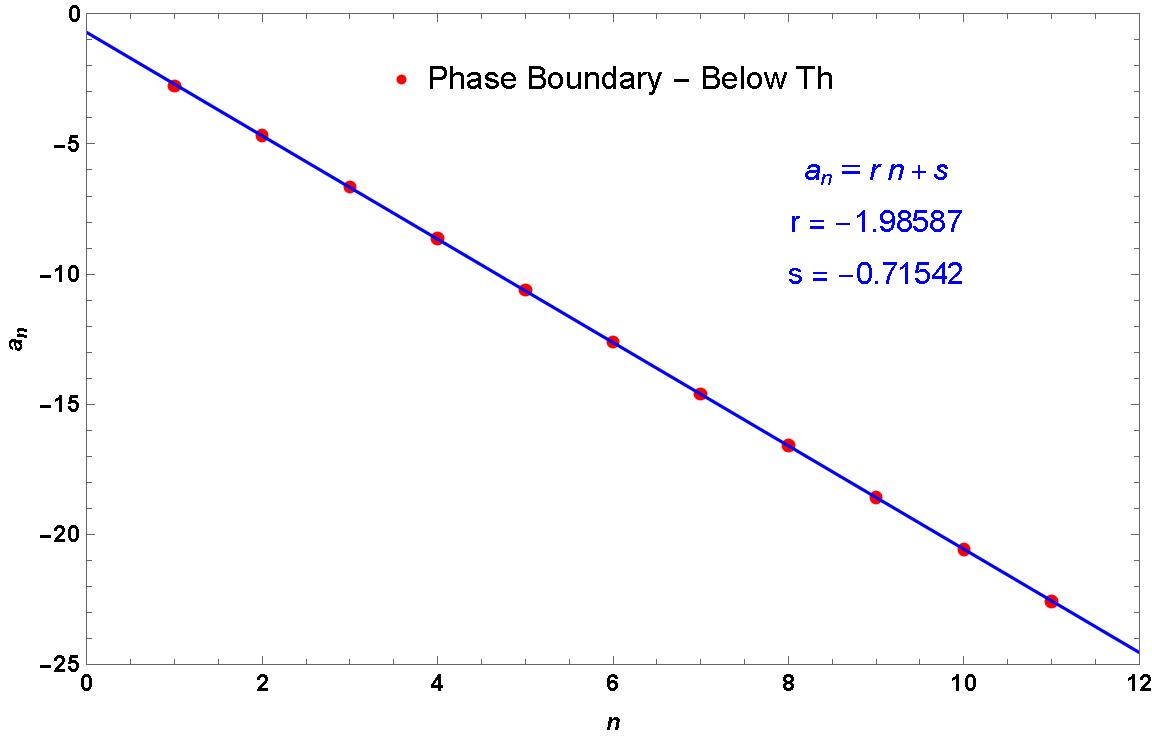}
			\caption{}
			\label{blanczosphasebdsbel}
		\end{subfigure}
		\caption{The above plots show the behavior of the Lanczos coefficients at the phase boundary below the threshold value. Plot [a]. shows Lanczos coefficient $b_n$ increases linearly in $n$ and [b]. shows Lanczos coefficient $a_n$ decreases with $n$. The red dots are the numerical data points, and the blue line represents the best fit, determined by specific values of the parameter sets (p, q) and (r, s).}
		\label{lanczospb}
  \end{figure}

In this section, we briefly outline the numerical technique utilized to compute K-Complexity. In order to do this we first determine the Lanczos coefficients numerically from the respective auto-correlations utilizing either the Toda chain method \cite{Dymarsky:2019elm} (also see appendix of \cite{Kundu:2023hbk}) or a modified version of moment method for non-zero $a_n$s described in \cite{VMR:2008vsg}. Since the Lanczos coefficients grow linearly up to some large-n we extrapolate the results and utilize them to obtain $\phi_n$s through the method described below. In all the cases we have considered, we find that the Lanczos grow linearly in n as expected in a QFT. We have given plots for some of the examples in Fig.[\ref{lanczosheating}] and Fig.[\ref{lanczospb}].

Determining the $\phi_n$s by directly solving the recursion equation in \eqref{phin} is not very tenable numerically. Hence, we resort to an interesting technique described in \cite{Bhattacharya:2023zqt}. The method involves re-expressing the recursion equation in \eqref{phin} as a matrix differential equation expressed below 
\begin{align}
    \frac{d \Phi(t)}{d t}={\tilde{\cal L}}\Phi(t)
\end{align}    
where $\tilde{{\cal L}}$ is given by
\begin{align}
	\tilde{\mathcal{L}}=\left(\begin{array}{cccccc}
		ia_0 & -b_1 & & & & 0 \\
		b_1 & ia_1 & -b_2 & & & \\
		&b_2 & ia_2 & \ddots & & \\
		& & \ddots & \ddots &-b_m & \\
		& & & b_m & ia_m & \ddots \\
		0 & & & & \ddots & \ddots
	\end{array}\right) .
\end{align}
and $\Phi$ is a column vector made up of the amplitudes $\phi_n$
\begin{align}
 \Phi(t)=   \left(\begin{array}{c}
\phi_0(t) \\
\phi_1(t) \\
\phi_2(t) \\
\vdots \\
\phi_{n_{max}-2}(t) \\
\phi_{n_{max}-1}(t)
\end{array}\right)
\end{align}
Note the norm of the vector $\Phi(t)$ gives the sum of the probabilities. The complexity on the other hand is given by the norm of the following column vector

\begin{align}
\sqrt{\mathcal{K}(t)}= \left(\begin{array}{c}
0 \\
\sqrt{1} \phi_1(t) \\
\sqrt{2} \phi_2(t) \\
\vdots \\
\sqrt{n_{max}-2}\ \phi_{n_{max}-2}(t) \\
\sqrt{n_{max}-1}\ \phi_{n_{max}-1}(t)
\end{array}\right)
\end{align}

We then solved the matrix differential equation numerically to obtain the results we have shown in our article. Although this method is more suitable for quantum systems with finite-dimensional Krylov space, it turns out to be very efficient in accurately reproducing the dynamics of the infinite-dimensional case up to a certain $t_{max}$ for a large enough matrix (For the results we have obtained we have gone upto a matrix of dimensions as high as 2000 $\times$ 2000). This $t_{max}$ can be determined approximately by comparing $\phi_0(t)$ obtained from numerical computation with the exact auto-correlation we derived. We have also verified that this method accurately reproduces the known results.


		\end{document}